\pdfoutput=1
\documentclass[aps,prd,twocolumn,superscriptaddress,preprintnumbers,floatfix,nofootinbib]{revtex4-1}

\usepackage{graphicx}
\usepackage{amsmath}
\usepackage[caption=false]{subfig}
\usepackage{siunitx}
\usepackage{placeins}
\usepackage{color}
\usepackage{hhline}
\usepackage{standalone}
\usepackage{dcolumn}
\usepackage{tensor}
\usepackage{bm}
\usepackage{microtype}
\usepackage{etoolbox}
\usepackage{xspace}
\usepackage{amssymb}
\usepackage{mathrsfs}
\usepackage{accents}
\usepackage[normalem]{ulem}
\usepackage[dvipsnames]{xcolor}
\usepackage[colorlinks,urlcolor=NavyBlue,citecolor=NavyBlue,linkcolor=NavyBlue,pdfusetitle]{hyperref}
\usepackage[all]{hypcap}
\usepackage[inline]{enumitem}
\usepackage[utf8]{inputenc}
\usepackage{lipsum}
\usepackage{booktabs}
\usepackage{wasysym}

\usepackage{array}
\newcolumntype{L}[1]{>{\raggedright\let\newline\\\arraybackslash\hspace{0pt}}m{#1}}
\newcolumntype{C}[1]{>{\centering\let\newline\\\arraybackslash\hspace{0pt}}m{#1}}
\newcolumntype{R}[1]{>{\raggedleft\let\newline\\\arraybackslash\hspace{0pt}}m{#1}}

\newcommand{\beq}{\begin{equation}}
\newcommand{\eeq}{\end{equation}}

\newtoggle{commentsoff}
\togglefalse{commentsoff}
\ifdefined\nocomments
    \toggletrue{commentsoff}
\fi

\iftoggle{commentsoff}{
  \newcommand*{\SB}[1]{}
  \newcommand*{\MI}[1]{}
  \newcommand*{\sv}[1]{}
  \newcommand*{\VV}[1]{}
  \newcommand*{\comment}[1]{}
  
  \newcommand*{\todo}[1]{}
  \newcommand*{\warn}[1]{}

}{
  \newcommand*{\MI}[1]{{\color{magenta} [{\bf MAX}: #1]}}
  \newcommand*{\SB}[1]{{\color{RedOrange} [{\bf SYLVIA}: #1]}}
  \newcommand*{\sv}[1]{\textcolor{ForestGreen}{[\textbf{SALVO}: #1]}}
  \newcommand*{\VV}[1]{\textcolor{Purple}{\textbf{VIJAY}: #1}}
  
  \newcommand*{\comment}[1]{{\color{blue} [{\bf NOTE}: #1]}}
  \newcommand*{\warn}[1]{{\color{red} [{\bf WARNING}: #1]}}
  \newcommand*{\todo}[1]{{\color{red} [{\bf TODO}: #1]}}

}

\graphicspath{{./}}

\newcommand{\MKI}{\affiliation{Department of Physics and Kavli Institute for Astrophysics and Space Research, Massachusetts Institute of Technology, 77 Massachusetts Ave, Cambridge, MA 02139, USA}}

\newcommand{\chieff}{\ensuremath{\chi_{\mathrm{eff}}}\xspace}

\begin{document}

\title{Measuring the spins of heavy binary black holes}

\author{Sylvia Biscoveanu}
\email[]{sbisco@mit.edu}
\affiliation{
LIGO Laboratory, Massachusetts Institute of Technology, Cambridge, Massachusetts 02139, USA
}%
\MKI

\author{Maximiliano Isi}
\email[]{maxisi@mit.edu}
\thanks{NHFP Einstein fellow}
\affiliation{
LIGO Laboratory, Massachusetts Institute of Technology, Cambridge, Massachusetts 02139, USA
}%
\MKI

\author{Vijay Varma}
\email{vvarma@cornell.edu}
\thanks{Klarman fellow}
\affiliation{Department of Physics, and Cornell Center for Astrophysics and
Planetary Science, Cornell University, Ithaca, New York 14853, USA}

\author{Salvatore Vitale}
\email{svitale@mit.edu}
\affiliation{
LIGO Laboratory, Massachusetts Institute of Technology, Cambridge, Massachusetts 02139, USA
}
\MKI
\hypersetup{pdfauthor={Biscoveanu, Isi, Varma, Vitale}}

\date{\today}

\begin{abstract}
An accurate and precise measurement of the spins of individual merging black holes is required to understand their origin.
While previous studies have indicated that most of the spin information comes from the inspiral part of the signal, the informative spin measurement of the heavy binary black hole system GW190521 suggests that the merger and ringdown can contribute significantly to the spin constraints for such massive systems. We perform a systematic study into the measurability of the spin parameters of individual heavy binary black hole mergers using a numerical relativity surrogate waveform model including the effects of both spin-induced precession and higher-order modes.  We find that the spin measurements are driven by the merger and ringdown parts of the signal for GW190521-like systems, but the uncertainty in the measurement increases with the total mass of the system. We are able to place meaningful constraints on the spin parameters even for systems observed at moderate signal-to-noise ratios, but the measurability depends on the exact six-dimensional spin configuration of the system. Finally, we find that the azimuthal angle between the in-plane projections of the component spin vectors at a given reference frequency cannot be well-measured for most of our simulated configurations even for signals observed with high signal-to-noise ratios.
\end{abstract}

\maketitle

\section{Introduction}
Gravitational-wave observations offer a unique way to directly measure the
masses and spins of astrophysical black holes, providing insights into their
formation channels~\cite{Gerosa:2013laa,Vitale:2015tea, Stevenson:2015bqa,Rodriguez:2016vmx,Stevenson:2017dlk, Talbot:2017yur, Fishbach:2017zga, Farr:2017gtv, Zevin:2017evb, Barrett:2017fcw, Taylor:2018iat, Sedda:2018nxm, Roulet:2018jbe, Wysocki_2019, Sedda:2020vwo, Baibhav:2020xdf, Kimball:2020opk, Zevin:2020gbd}. 
The latest observing run of the advanced
LIGO~\cite{TheLIGOScientific:2014jea} and Virgo~\cite{TheVirgo:2014hva}
gravitational-wave interferometers resulted in the discovery of tens of new binary black hole
systems~\cite{Abbott:2020niy} leading to the best constraints on the
population-level mass and spin distributions of these objects to
date~\cite{Abbott:2020gyp}. Reliable measurements of the masses and spins of
individual black holes are crucial ingredients for population analyses and of
particular interest for exceptional events. 

GW190521 is one such exceptional event, representing the first direct detection of an intermediate-mass
black hole~\cite{Abbott:2020tfl}. The inferred component masses of this system
are above the theoretical upper limit for black holes formed directly from
stellar collapse due to the effects of pair-instability
supernovae~\cite{Heger:2001cd, Ozel:2010su, Belczynski:2016jno,
Marchant:2016wow, Woosley:2016hmi} (although see \cite{Belczynski:2020bca,
Fishbach:2020qag, Nitz:2020mga}). Together with some evidence for
general-relativistic spin-induced precession, this
suggests that this binary could have formed in a dynamical
environment~\cite{PortegiesZwart:1999nm, PortegiesZwart:2002iks, Miller:2001ez,
OLeary:2005vqo, Samsing:2013kua, Morscher:2014doa, Fragione:2018vty} as the
result of a hierarchical merger~\cite{Rodriguez:2017pec, Gerosa:2017kvu,
Rodriguez:2019huv, Bouffanais:2019nrw, Fragione:2020aki}. Previous works have
claimed evidence for eccentricity in this
system~\cite{Romero-Shaw:2020thy, Gayathri:2020coq, CalderonBustillo:2020srq},
another signature of dynamical formation.

Intermediate-mass black holes, with masses approximately
$10^{2}-10^{5}~M_{\odot}$, occupy the mass range in between stellar-mass and
supermassive black holes, lying
above the putative pair-instability mass gap described above and potentially
serving as the seeds for the supermassive black holes at the centers of
galaxies~\cite{Ebisuzaki:2001qm, Miller:2003sc, Mezcua:2017npy,
Greene:2019vlv}. Such objects can form directly in the early Universe via the
collapse of massive Population III stars~\cite{Fryer:2000my, Heger:2002by,
Spera:2017fyx, Liu:2021jdz} or low-angular-momentum gas clouds~\cite{Loeb:1994wv,
Bromm:2002hb, Lodato:2006hw, Begelman:2006db}, or dynamically in dense
environments via runaway collisions~\cite{PortegiesZwart:2002iks,
PortegiesZwart:2004ggg, AtakanGurkan:2003hm} or hierarchical
mergers~\cite{Miller:2001ez, OLeary:2005vqo, 2015MNRAS.454.3150G}, as suggested
above. Given the unique formation channels of binaries involving these objects
and the fact that the sensitivity of ground-based gravitational-wave detectors
increases with the total mass of the binary (up to a few hundred solar masses)~\cite{Fishbach:2017zga,
Ezquiaga:2020tns}, such systems are of particular interest.

Because gravitational-wave frequency scales inversely with the total mass, heavy binaries merge at lower frequencies, leaving few orbital cycles in the sensitive band of ground-based
detectors.  This would imply that our ability to constrain the spin parameters, which is traditionally thought to be driven by the inspiral part of the waveform, is limited for such heavy systems.
However, the informative spin measurement and evidence for precession found in GW190521 seem to indicate that meaningful inferences can be made for the spins of heavy binary black holes (BBHs). In this work, we seek to systematically investigate how well spins can be measured for such heavy systems.

To leading post-Newtonian order, the influence of the spins can be parameterized in terms of a single mass-weighted effective aligned spin
parameter, \chieff~\cite{Damour:2001, Ajith:2009bn, Ajith:2011ec, Santamaria:2010yb,
Purrer:2013ojf}:
\begin{align}
\chi_{\mathrm{eff}} &= \frac{\chi_{1}\cos{\theta_{1}} + \chi_{2}q\cos{\theta_{2}}}{1+q}, \label{eq:chi_eff}
\end{align}
where $q\equiv m_{2}/m_{1},\ m_{2} \leq m_{1}$ is the binary mass ratio, $\chi_{i}$ is the component spin magnitude with $i =1,2$, and $\theta_{i}$ is the component spin tilt relative to the orbital angular momentum.
For binaries where the component spins are aligned or antialigned with the orbital angular momentum, the spin modifies the inspiral rate relative to the equivalent
non-spinning system, but the waveforms are otherwise morphologically
similar~\cite{Campanelli:2006uy}. 

However, when the spins of the black holes are
misaligned to the orbital angular momentum, the orbital plane and
the individual spins precess about the total angular
momentum~\cite{Apostolatos:1994mx, Kidder:1995zr}. The waveforms for these
systems are morphologically richer, leading to characteristic modulations in the
gravitational-wave amplitude and phase. Even in this
generic case, however, the average influence of the remaining four spin degrees
of freedom can be reduced to a single effective precessing spin parameter,
$\chi_{p}$\footnote{See \cite{Thomas:2020uqj, Gerosa:2020aiw} for alternative
    definitions of an effective precessing spin parameter that more robustly
capture the effects of higher-order modes and variations occurring on the
precessional timescale, respectively.}~\cite{Schmidt:2014iyl}:
\begin{align}
\chi_{p} &= \max{\left(\chi_{1}\sin{\theta_{1}}, \left(\frac{4q+3}{4+3q}\right)q\chi_{2}\sin{\theta_{2}}\right)}. \label{eq:chi_p}
\end{align}

Because these
parameters encode the leading order effect of spins on the inspiral
phasing, \chieff and
$\chi_{p}$ are typically much better measured than the individual component
spins for signals dominated by the
inspiral~\cite{Vitale:2016avz, Shaik:2019dym}.
Most
previous studies investigating the measurability of spin parameters with
ground-based gravitational-wave detectors have focused on systems in this
inspiral-dominated regime~\cite{vanderSluys:2007st, Raymond:2009cv, Cho:2012ed,
OShaughnessy:2014shr, Vitale:2014mka, Ghosh:2015jra, Chatziioannou:2018wqx,
Pratten:2020igi, Green:2020ptm}. Such work finds that the
secondary spin is particularly hard to measure~\cite{Raymond:2009cv}, as are
high component spins when \chieff is small~\cite{Chatziioannou:2018wqx}.
However, unequal masses significantly improve the resolution of the spin
parameters~\cite{Vitale:2014mka, Pankow:2016udj}, particularly when
higher-order multipoles are included in the waveform model~\cite{Cho:2012ed,
OShaughnessy:2014shr, Pratten:2020igi}.  Precession effects can further serve
to break degeneracies and improve spin constraints~\cite{vanderSluys:2007st,
Cho:2012ed, OShaughnessy:2014shr, Pratten:2020igi}.

While \chieff and $\chi_{p}$ encapsulate the spin effects for the inspiral part
of the waveform, heavier binary black hole systems will have significant
contributions to the detectable signal from the merger and ringdown. The morphology of the latter is
determined by the mass and spin of the final black hole. Most previous studies
investigating the ability to
characterize such heavy systems have focused on
nonspinning~\cite{Graff:2015bba, Haster:2015cnn, Veitch:2015ela} or
aligned-spin~\cite{Purrer:2015nkh, Shaik:2019dym, Mehta:2021fgz} BBHs. Ref.
\cite{Vitale:2016avz} included precessing systems in their analysis, but used a
waveform model that only includes the contribution from the dominant $\ell =
|m| =2$ mode to the signal. This phenomenological waveform model,
IMRPhenomPv2~\cite{Hannam:2013oca, Husa:2015iqa, Khan:2015jqa}, approximates
the effects of precession in two important ways: (i) only
four spin degrees of freedom are included, and the azimuthal spin angles
are ignored; and (ii) this model is not informed by numerical relativity (NR)
simulations of precessing binaries, which are critical to accurately model the
merger process.  Instead, an equivalent nonprecessing waveform is ``twisted up''
to account for some effects of orbital precession~\cite{Schmidt:2012rh}. Furthermore, IMRPhenomPv2 and other such phenomenological waveform models rely on heuristic recipes for modeling the ringdown of the remnant black hole and attaching it to the rest of the
waveform. Accurate ringdown modeling is especially important for high-mass systems since this part of the waveform contributes significantly to the detectability of the signal. 

Considering the importance of precession, ringdown modeling, and higher-order modes 
for heavy binary black hole systems~\cite{Shaik:2019dym}, we seek to expand upon
previous studies to determine how well the spins can be measured for such
systems using a numerical relativity surrogate waveform that includes the
effects of the full six-dimensional spin degrees of freedom and higher
harmonics, NRSur7dq4~\cite{Varma:2019csw}. Trained directly on precessing
NR simulations, this model captures the effects of
precession at an accuracy level comparable to the simulations themselves. 

The rest of the paper is organized as follows.
In Section~\ref{sec:pe}, we describe the
Bayesian inference formalism we employ for our analysis.
Section~\ref{sec:high_chip} presents an investigation into the measurability of
precession in highly precessing systems comparable to GW190521. A systematic
study of the measurability of the spin degrees of freedom is presented in
Section~\ref{sec:bias_vs_mass} for the component and effective spins, 
and in Section~\ref{sec:phases} for the
azimuthal angles. We find that the spin information is driven by the merger and ringdown parts of the signal for GW190521-like systems, but that the measurability of the spin parameters depends sensitively on the exact six-dimensional spin configuration of the binary.

\section{Parameter Estimation}
\label{sec:pe}

\begin{figure*}
\centering
\includegraphics[width=0.85\columnwidth]{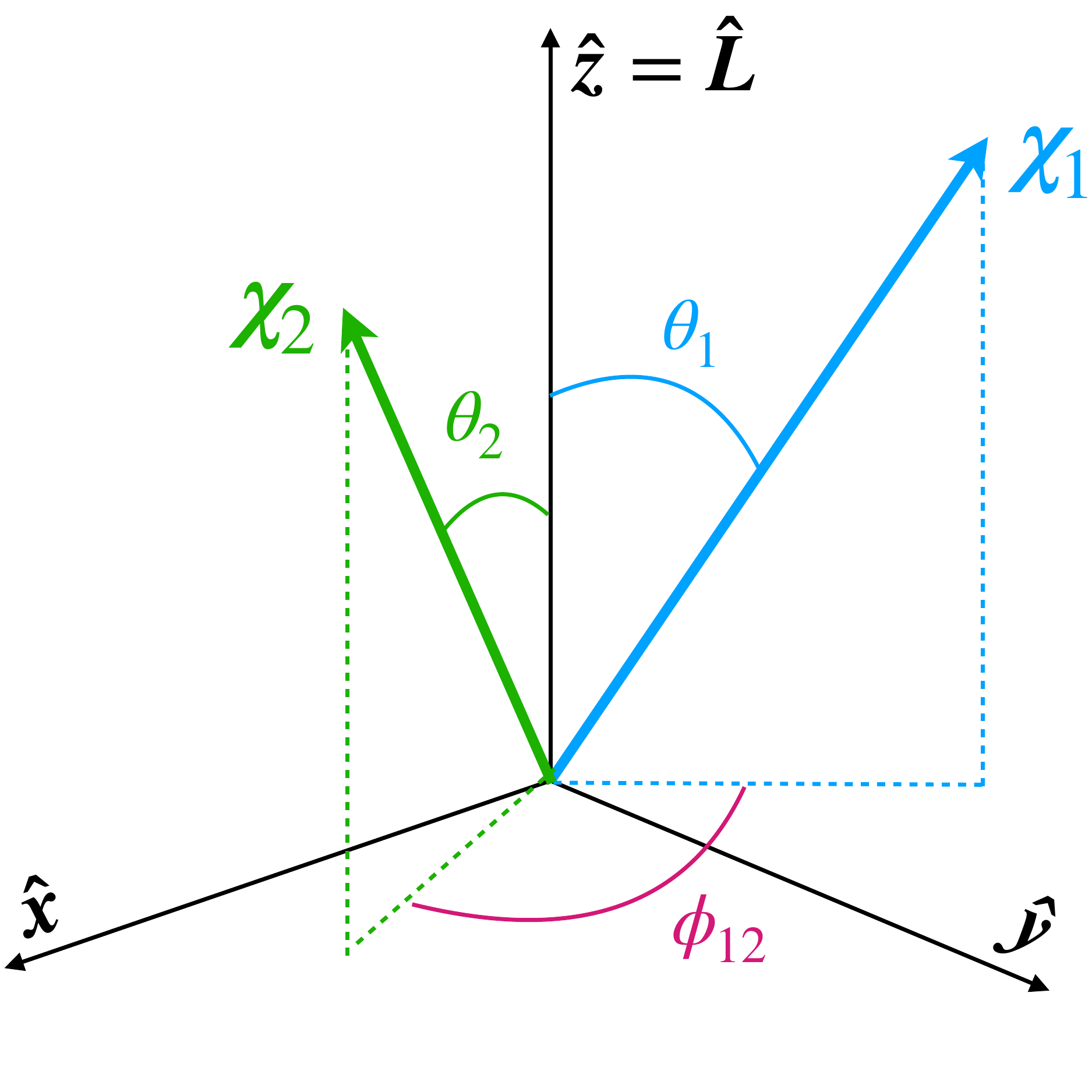}
\includegraphics[width=0.85\columnwidth]{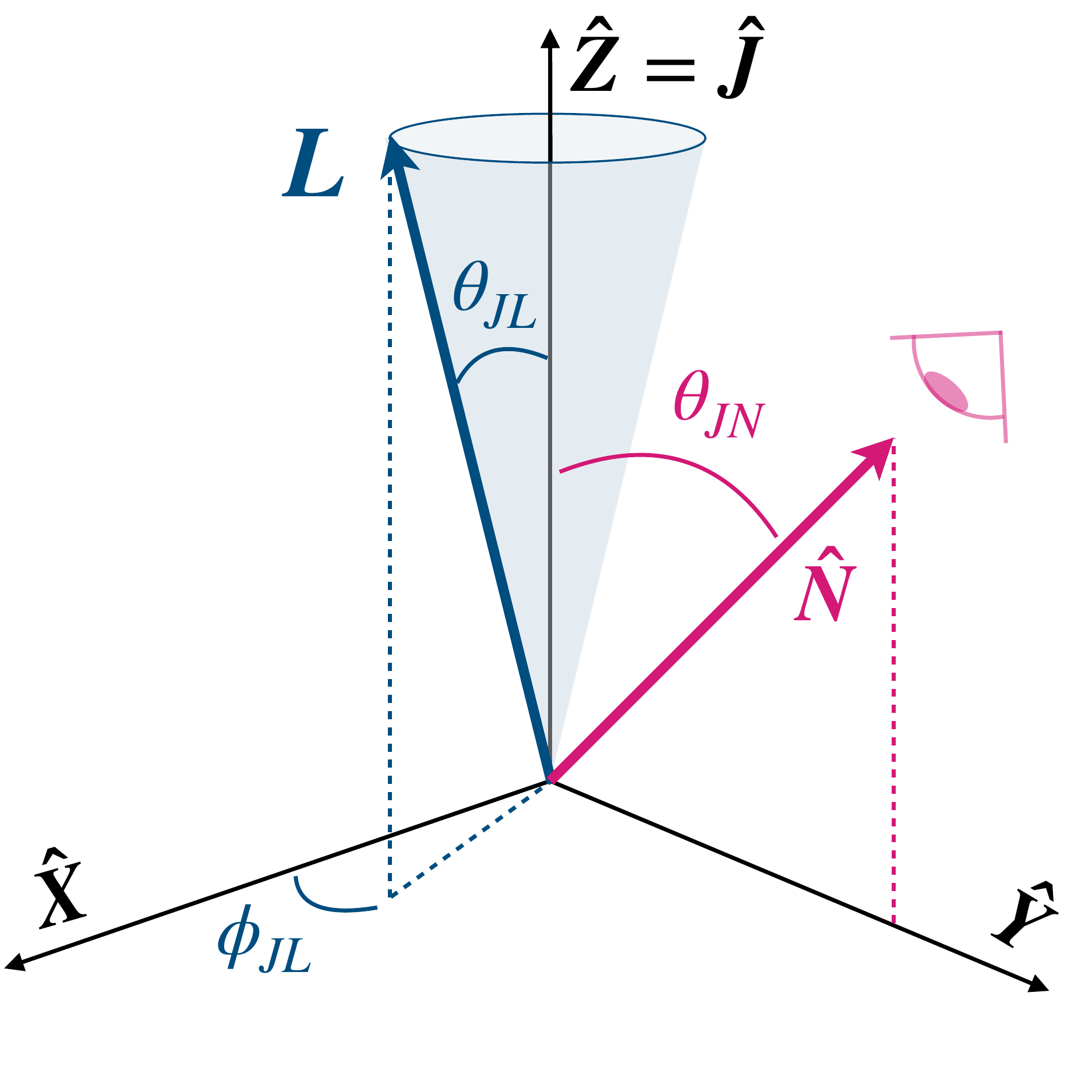}
\caption{\textit{Left:} Diagram of the component spin vectors relative to the orbital angular momentum pointing along the $\hat{z}$ axis. $\phi_{12}$ is the azimuthal angle between the projections of the component spins onto the orbital plane. \textit{Right:} Diagram demonstrating the definition of $\phi_{JL}$, which is the azimuthal angle between the $\hat{X}$ axis and the projection of the orbital angular momentum onto the $\hat{X}-\hat{Y}$ plane in the frame where the total angular momentum points along the $\hat{Z}$ axis and the line-of-sight vector, $\hat{N}$, lies in the $\hat{Y}-\hat{Z}$ plane.
}
\label{fig:schematic}
\end{figure*}

A quasicircular binary black hole merger is completely characterized by 15
parameters: eight intrinsic parameters comprising of the
masses and spin vectors for each component black hole, and seven extrinsic
parameters. The extrinsic parameters include the sky location $(\alpha,
\delta)$, the luminosity distance to the source, $d_{L}$, the inclination angle
between the total angular momentum and the line of sight of the observer,
$\theta_{JN}$, the polarization angle $\psi$, and the time and orbital phase
$\phi$ at coalescence. The spin degrees of freedom are typically parameterized
in terms of the component magnitudes, $\chi_{i}$, tilt angles relative to the
orbital angular momentum $\theta_{i}$, and two azimuthal angles, $\phi_{12}$
and $\phi_{JL}$, defined in the diagrams in Fig.~\ref{fig:schematic}. The component black holes are usually sorted according to
their masses, with $\chi_{1}$ indicating the spin magnitude of the more massive
black hole, but they can also be sorted according to their spins, with subscripts $A, B$~\cite{Biscoveanu:2020are}.

In order to obtain estimates of these 15 binary parameters from
gravitational-wave strain data, we employ the framework of Bayesian inference.
The posterior probability distribution for the binary parameters
$\boldsymbol{\theta}$ given the observation of data $d$ is:
\begin{align}
p(\boldsymbol{\theta} | d) \propto \mathcal{L}(d | \boldsymbol{\theta})\pi(\boldsymbol{\theta}).
\label{eq:post}
\end{align}
The likelihood of observing the data $d$ is~\cite{Veitch:2014wba, Romano:2016dpx} 
\begin{align}
\mathcal{L}(d | \boldsymbol{\theta}) \propto \exp{\left(-\sum_{k} \frac{2 | d_{k} - h_{k}(\boldsymbol{\theta}) |^{2}}{TS_{k}}\right)},
\label{eq:likelihood}
\end{align}
where $h(\boldsymbol{\theta})$ represents the gravitational waveform which
depends on the binary parameters, $T$ is the duration of the analyzed data, $S$
is the power spectral density (PSD) characterizing the noise in the
interferometer, and the subscript $k$ indicates the frequency dependence of the
data, waveform, and PSD. The prior probability distributions for the binary
parameters are represented by
$\pi(\boldsymbol{\theta})$. 

We employ the formalism described above to perform parameter estimation for
high-mass binary black hole merger simulations. Unless otherwise stated, the
following settings are used for all the simulations described in this
manuscript. We use a three-detector network of the advanced
LIGO~\cite{TheLIGOScientific:2014jea} and Virgo~\cite{TheVirgo:2014hva}
interferometers operating at design sensitivity and do not add Gaussian noise
to the simulated data in each detector so that the analysis is performed in zero-noise~\cite{Vallisneri:2007ev}. 
We analyze $8~\mathrm{s}$ segments of
data sampled at a rate of $2048~\mathrm{Hz}$. The
signals are simulated and recovered with the NRSur7dq4
waveform~\cite{Varma:2019csw} for $h(\boldsymbol{\theta})$, and samples from
the posterior distribution given in Eq.~\eqref{eq:post} are obtained using the
LALInference software~\cite{Veitch:2014wba}. 

NRSur7dq4 is trained on numerical relativity
simulations with mass ratios $q\geq1/4$ and spin magnitudes
$\chi_1,\chi_2\leq0.8$, but extrapolates reasonably well to $q\leq1/6$ and
$\chi_1,\chi_2=1$~\footnote{While the number of precessing NR simulations with spins $\gtrsim 0.8$ are limited, we have checked that for the $\sim 30$ such available waveforms, the surrogate agrees with NR to within NR resolution errors. We have also verified that parameter estimation results within the training region are unaffected when using priors extending into the extrapolation region relative to those restricted to the training region.}. We therefore allow our priors to extend to $q_{\min}=1/5$ and $\chi_{\max}=0.99$.
Unless otherwise stated, the reference frequency at which
the spin parameters are specified is $25~\mathrm{Hz}$, and this is also the
minimum frequency included in the likelihood.
While advanced LIGO and Virgo are sensitive to lower frequencies,
NRSur7dq4 only includes about $\sim 20$ orbits before
merger~\cite{Varma:2019csw}, which can lead to the signal abruptly starting
within the detector band for small total masses or small mass ratios. To avoid
this, we use a conservative minimum frequency of $25~\mathrm{Hz}$ so that NRSur7dq4
waveforms can be successfully generated for the smallest total mass
($M_{\mathrm{tot}} = 70~M_{\odot}$) and mass ratio we allow in our
prior. We use priors that are uniform in
the component masses, spin magnitudes, azimuthal angles and
$\cos{\theta_{i}}$,\footnote{See \cite{Vitale:2017cfs, Huang:2020ysn, Zevin:2020gxf} for discussions on the effects of prior choices on spin measurements.} and $\propto
d_{L}^{2}$ in luminosity distance. Standard priors are used for all other
parameters~\cite{Veitch:2014wba}.

\section{Measurability of spins in GW190521-like systems}
\label{sec:high_chip}
GW190521 is the heaviest binary black hole system detected in gravitational waves to date, and its $\chi_{p}$ posterior shows a noticeable deviation from the prior, peaking at around $\chi_{p}\approx
0.7$~\cite{Abbott:2020tfl}. This suggests that general-relativistic spin-induced 
precession is measurable in high-mass signals with moderate
signal-to-noise ratios (SNRs).  We therefore seek to
verify this measurement using three simulated systems of varying inclination
angle and mass ratio, $q$, (and hence varying \chieff) in order to determine how the
measurability of $\chi_{p}$ depends on these parameters. The distance for each
system is adjusted so that the network optimal SNR is fixed to
$\rho_{\mathrm{opt}}^{\mathrm{net}} = 15$ to match the SNR of GW190521. The
true value of $\chi_{p} = 0.90$ for all three simulations, and the remaining
parameters which are kept the same are presented in Table~\ref{tab:chip}.
 For these systems we use
minimum and reference frequencies of $11~\mathrm{Hz}$, as was done in the
analysis of GW190521~\cite{Abbott:2020tfl}. The three choices of inclination
and mass ratio are: 
\begin{enumerate}
\item Nearly edge-on, nearly equal-mass: $\theta_{JN}=1.4,\ q=0.9,\ \chieff = 0.074$
\item Nearly face-on, nearly equal-mass: $\theta_{JN}=0.5,\ q=0.9,\ \chieff = 0.074$
\item Nearly edge-on, more unequal masses: $\theta_{JN}=1.4,\ q=0.7,\ \chieff = 0.067$
\end{enumerate}

\begin{table}
    \caption{True values for the parameters which are kept the same for all
    three simulations with SNR=15 shown in Fig.~\ref{fig:chip_corner}.}
\label{tab:chip}
\begin{ruledtabular}
\begin{tabular}{l l l}
    Parameter & Symbol & Value\\
    \midrule
    Detector-frame total mass & $M_{\mathrm{tot}}$ & $270~M_{\odot}$ \\
    Primary spin magnitude & $\chi_{1}$  & 0.9  \\
    Secondary spin magnitude & $\chi_{2}$  & 0.8 \\
    Primary tilt\footnote{All angles in radians.} & $\theta_{1}$ & 1.55 \\
    Secondary tilt & $\theta_{2}$ & 1.40 \\
    Azimuthal inter-spin angle & $\phi_{12}$ & 6.28 \\
    Azimuthal precession cone angle & $\phi_{JL}$ & 6.28 \\
    Effective precessing spin & $\chi_{p}$ & 0.90\\
    Coalescence phase & $\phi$ & 4.70 \\
    Polarization angle & $\psi$ & 0.28 \\
    Coalescence GPS time & $t_{c}$ & 1126259642 s\\
    Right ascension & $\alpha$ & 1.09 \\
    Declination & $\delta$ & 0.47 \\
\end{tabular}
\end{ruledtabular}
\end{table}

\begin{figure*}
\centering
\includegraphics[width=0.8\textwidth]{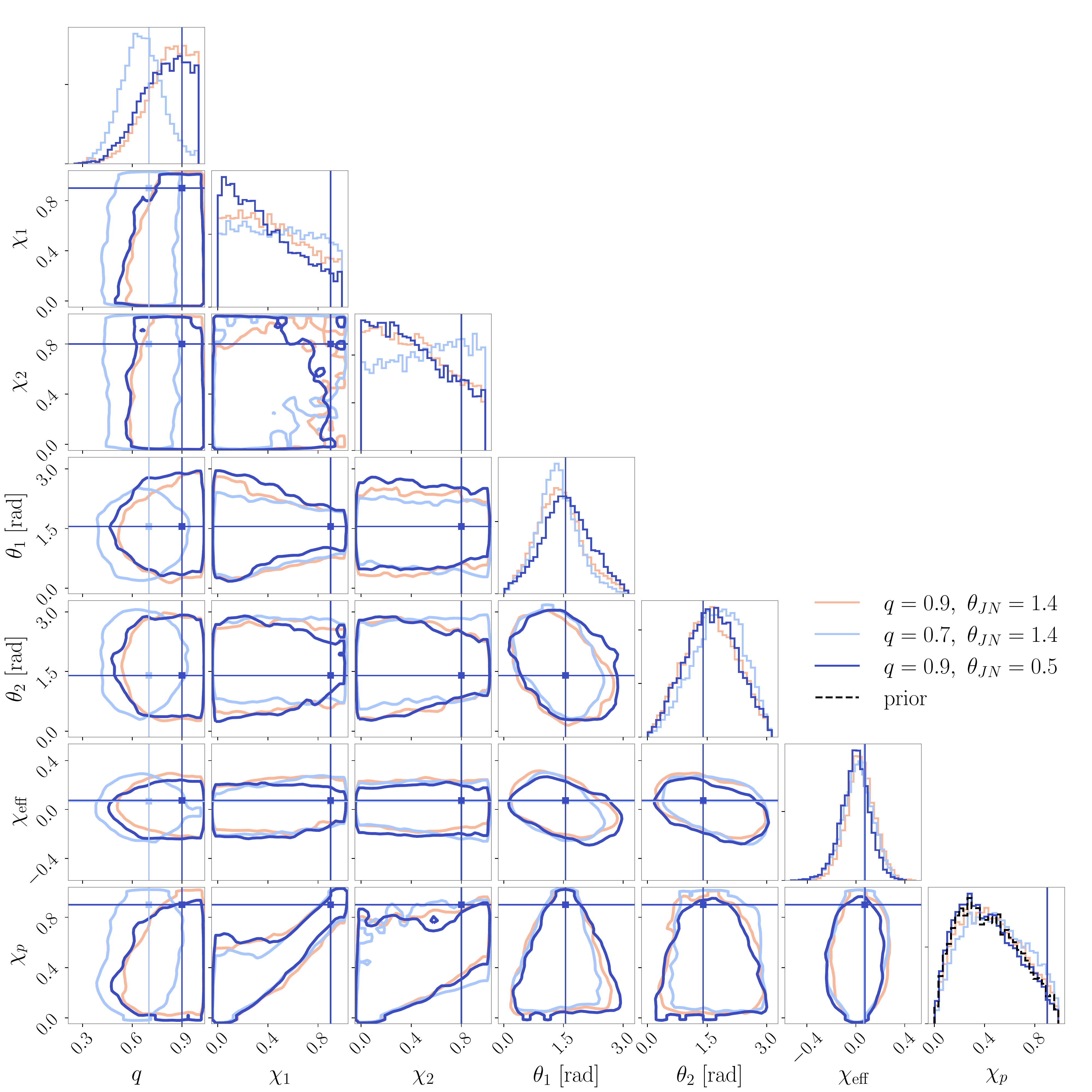}\qquad
\caption{Comparison corner plot for the three simulated systems with
$\chi_{p}=0.9$ at SNR=15 showing the posteriors for mass ratio, spin
magnitudes, \chieff, and $\chi_{p}$. The prior for $\chi_{p}$ is shown in the dashed black line.
}
\label{fig:chip_corner}
\end{figure*}

GW190521 is consistent with having been observed face-on~\cite{Abbott:2020tfl}, so it is most similar to simulation number 2 above. The results of our simulations are shown in Fig.~\ref{fig:chip_corner}. We are
unable to recover a significant measurement of $\chi_{p}$ matching the
measurement obtained for GW190521 with any of the three systems. In fact, the
posteriors for the component spin magnitudes prefer low values, peaking at
$\chi_{i}=0$ for both of the nearly-equal mass simulations. The spin magnitude
posteriors for the more unequal-mass system show less of a preference for low
spin magnitudes, although they only exhibit a minor deviation from the uniform
prior. The spin tilt posteriors are also not informative enough to drive the
measurement of $\chi_{p}$ to high values; while the true values of the tilt angles do
fall at the peak of the posterior, the posterior distributions are largely unconstrained relative to the prior. The resulting posterior for $\chi_{p}$ is also driven by the prior, shown in the dashed black line in Fig.~\ref{fig:chip_corner}. The fact that we are unable to recover an informative
posterior for $\chi_{p}$---even for the system observed nearly edge on, for
which the effects of precession should be most apparent---suggests that
obtaining a more informative measurement, such as the one for GW190521,
requires tuning to a specific configuration among the
spin degrees of freedom. It could also be an indication that the posterior for
GW190521 is driven to high values of $\chi_{p}$ due to the properties of the
detector noise at the time of the event. 

To investigate these possibilities, we conduct parameter estimation for a
simulated system with parameters specified by the maximum-likelihood point from
the publicly-released posterior samples obtained for GW190521 by LIGO-Virgo~\cite{data_release_GW190521, gwosc} using the
NRSur7dq4 waveform. This ensures that any parameter correlations that led to a more informative measurement of $\chi_{p}$ for GW190521
are included in our simulation. We add this BBH signal to ten different
realizations of Gaussian noise 
colored by the power spectral densities of the two
advanced LIGO detectors and advanced Virgo calculated for the segment of data
containing GW190521~\cite{data_release_GW190521}. We also use minimum and reference frequencies of $11~\mathrm{Hz}$ for analyzing this maximum-likelihood system. The binary parameters are given
in Table~\ref{tab:maxL}.

\begin{table}
   \caption{Maximum likelihood parameter values used for the simulated system drawn from the GW190521 posterior.}
\label{tab:maxL}
\begin{ruledtabular}
\begin{tabular}{l l l}
    Parameter & Symbol & Value\\
    \midrule
    Mass ratio & $q$ & 0.82 \\
    Detector-frame total mass & $M_{\mathrm{tot}}$ & $268.83~M_{\odot}$ \\
    Primary spin magnitude & $\chi_{1}$  & 0.09  \\
    Secondary spin magnitude & $\chi_{2}$  & 0.90 \\
    Primary tilt\footnote{All angles in radians.} & $\theta_{1}$ & 0.92 \\
    Secondary tilt & $\theta_{2}$ & 1.41 \\
    Azimuthal inter-spin angle & $\phi_{12}$ & 1.04 \\
    Azimuthal precession cone angle & $\phi_{JL}$ & 2.48 \\
    Effective aligned spin & \chieff & 0.094\\
    Effective precessing spin & $\chi_{p}$ & 0.70\\
    Coalescence phase & $\phi$ & 0.004 \\
    Polarization angle & $\psi$ & 2.38 \\
    Coalescence GPS time & $t_{c}$ & 1242442967.41 s\\
    Right ascension & $\alpha$ & 0.16 \\
    Declination & $\delta$ & -1.14 \\
    Luminosity distance & $d_{L}$ & 2941.03 Mpc \\
    Inclination angle & $\theta_{JN}$ & 0.95\\
\end{tabular}
\end{ruledtabular}
\end{table}

\begin{figure}
\centering
\includegraphics[width=0.5\textwidth]{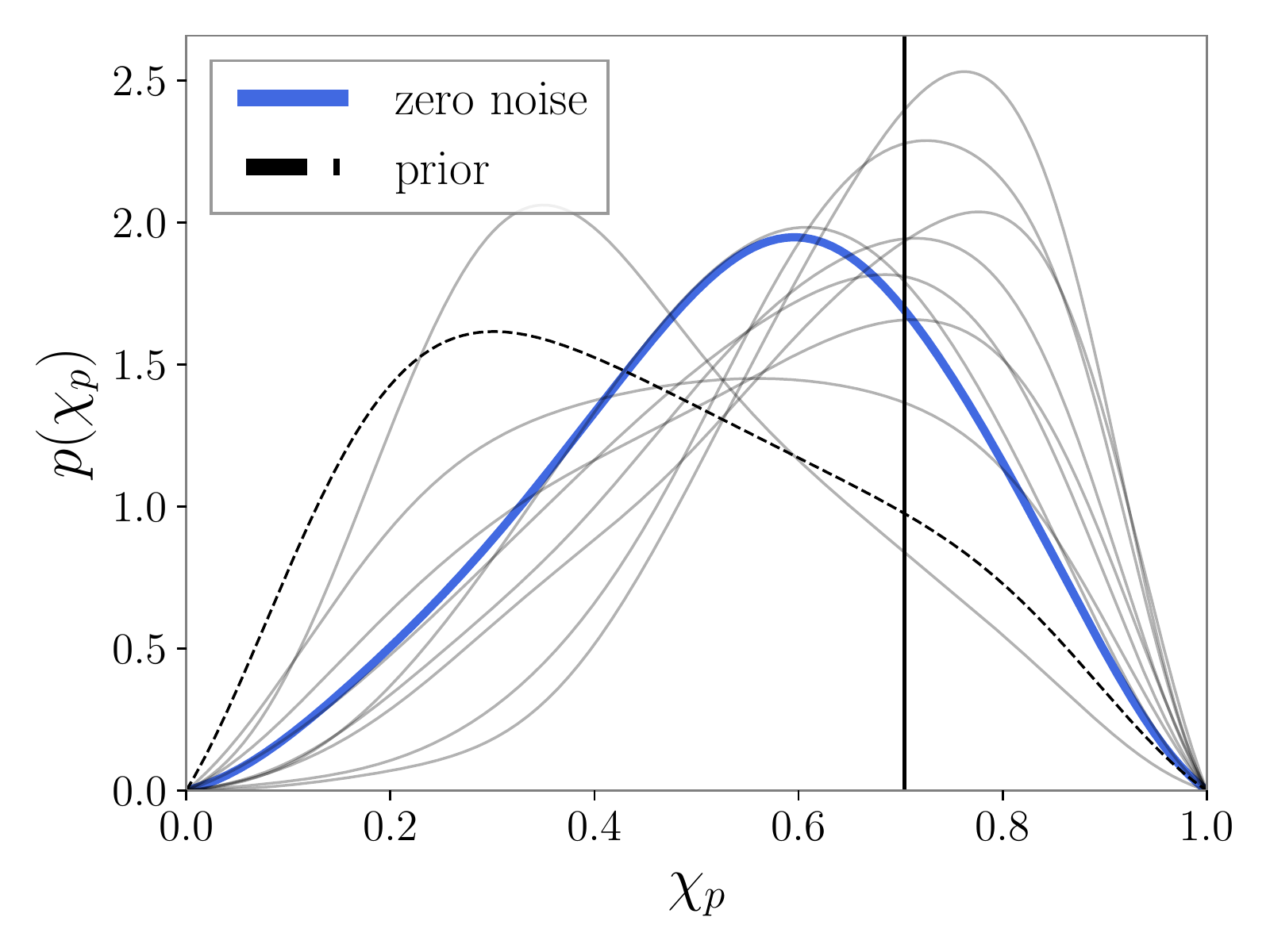}\qquad
\caption{Probability density for $\chi_{p}$ for the GW190521 maximum-likelihood
    simulation added to ten different realizations of Gaussian noise colored
    by the PSDs estimated for the
time around GW190521. The blue line shows
the result for the simulation without Gaussian noise. The prior is shown in the dashed black line.}
\label{fig:GW190521_high}
\end{figure}

The posterior probability distributions obtained for $\chi_{p}$ for the ten
different realizations of Gaussian noise are shown in
Fig.~\ref{fig:GW190521_high}, along with the posterior obtained in the absence
of Gaussian noise. Even though the true value of $\chi_{p} = 0.70$ is lower
than that of the previous three simulations, the measurement is much more
informative and peaks closer to the true value for the comparable zero-noise
case. While one of the noise realizations leads to a posterior peaked at low
values of $\chi_{p}$ and another returns a flatter posterior, these deviations are consistent with
Gaussianity. Because real interferometer data is known to
include non-Gaussian excursions, we repeat the experiment by adding the
maximum-likelihood system into ten different segments of real data from the
third observing run of advanced LIGO and Virgo instead of simulated Gaussian
noise. We find a similar spread in the $\chi_{p}$ posteriors to that shown in
Fig.~\ref{fig:GW190521_high}; further details can be found in
Appendix~\ref{ap:GW190521}.

These results indicate that even the spin degrees of freedom that are not usually measurable can have a significant effect on the detectability of precession. Only specific values of the spin angles yield a system
where $\chi_{p}$ is measured as well
as it was for GW190521. The $\chi_{p}$ posteriors shown in Fig.~\ref{fig:chip_corner} are qualitatively similar to those found in~\cite{Chatziioannou:2017tdw}, where it is suggested that high values of $\chi_{p}$ are more difficult to recover in systems with low values of \chieff.
However, the systems shown in Figs.~\ref{fig:chip_corner} and \ref{fig:GW190521_high}
have similar true values of \chieff, all $<0.1$, but very different posteriors
for $\chi_{p}$, suggesting that the measurability of $\chi_{p}$ depends on more than just \chieff.
While we have found that certain configurations can produce systems with informative $\chi_{p}$ posteriors, such configurations might be uncommon. If this is the case, more exotic, but also rare, explanations
could be invoked to explain the highly precessing nature of this event, such as
residual eccentricity at merger~\cite{Gayathri:2020coq, Romero-Shaw:2020thy},
which can be confused with precession when analyzed with waveforms assuming the
binary system is circular~\cite{Romero-Shaw:2020thy, CalderonBustillo:2020odh}.

\subsection{Effect of the cutoff frequency on the spin measurement}
\label{sec:fmax}
\begin{figure}
\centering
\includegraphics[width=0.5\textwidth]{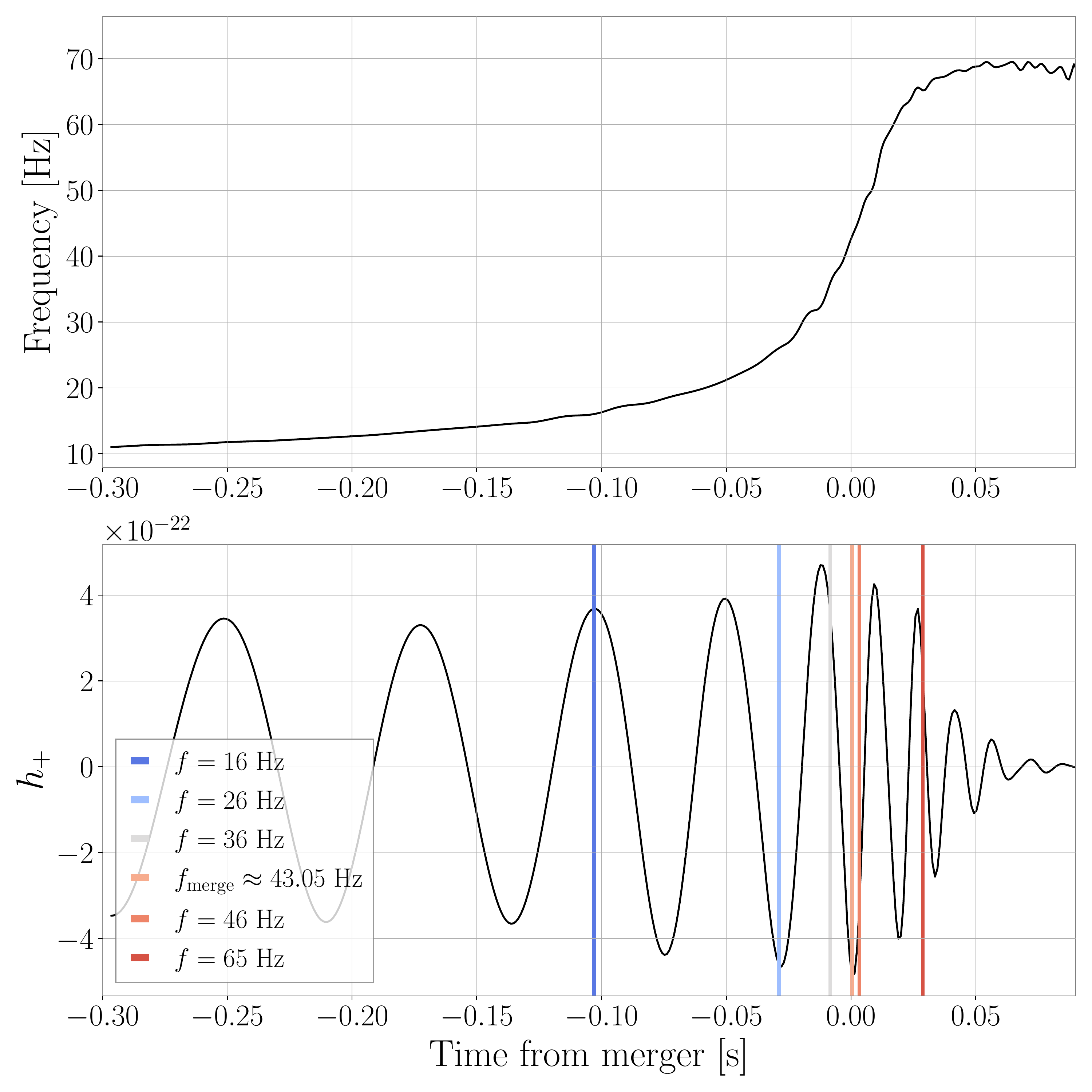}\qquad
\caption{\textit{Top panel:} Evolution of the gravitational-wave frequency in
the coprecessing frame as a function of time for the GW190521
maximum-likelihood simulation, with the merger time set to $t=0$.
\textit{Bottom panel:} Time-domain waveform with the varying maximum
frequencies we use in our analysis marked in time as vertical lines. The
frequency at the merger time is also shown.
} 
\label{fig:waveform}
\end{figure}

The information on the spin parameters in BBH sources is traditionally thought to come predominantly from the inspiral phasing, and yet in the previous section we have demonstrated that an informative measurement of $\chi_{p}$ can be made for a heavy GW190521-like system with few inspiral cycles in the sensitive band of the LIGO-Virgo detectors.
In order to determine which part of the frequency band drives the measurement
of the spin parameters for high-mass, highly-precessing systems, we perform a
study where the maximum frequency that is included in the likelihood in Eq.~\ref{eq:likelihood} for the
system in Table~\ref{tab:maxL} is incrementally increased. We begin with a maximum frequency of $16~\mathrm{Hz}$, so that the total analyzed bandwidth is only $5~\mathrm{Hz}$, but rescale the distance of the system so that the total SNR is kept constant at 15 for all the choices of $f_{\max}$. Gaussian noise is not added to these simulations. 
The time evolution of the gravitational-wave frequency in the coprecessing
frame~\cite{Varma:2019csw} is shown in the top panel of Fig.~\ref{fig:waveform}.  The
gravitational-wave frequency is calculated as $f_\mathrm{GW} =
\Omega_{\mathrm{orb}}/\pi$, where $\Omega_{\mathrm{orb}}$ is the time
derivative of the orbital phase as defined in Eq.~(3) of \cite{Varma:2019csw}.
The gravitational-wave amplitude peaks at
$f_{\mathrm{GW}}=43.05~\mathrm{Hz}$, which we treat as the merger frequency. The
analysis with the full band up to $512~\mathrm{Hz}$ is the same as in the dark
blue line in Fig.~\ref{fig:GW190521_high}.

\begin{figure*}[ht!]
\centering
\includegraphics[width=0.8\textwidth]{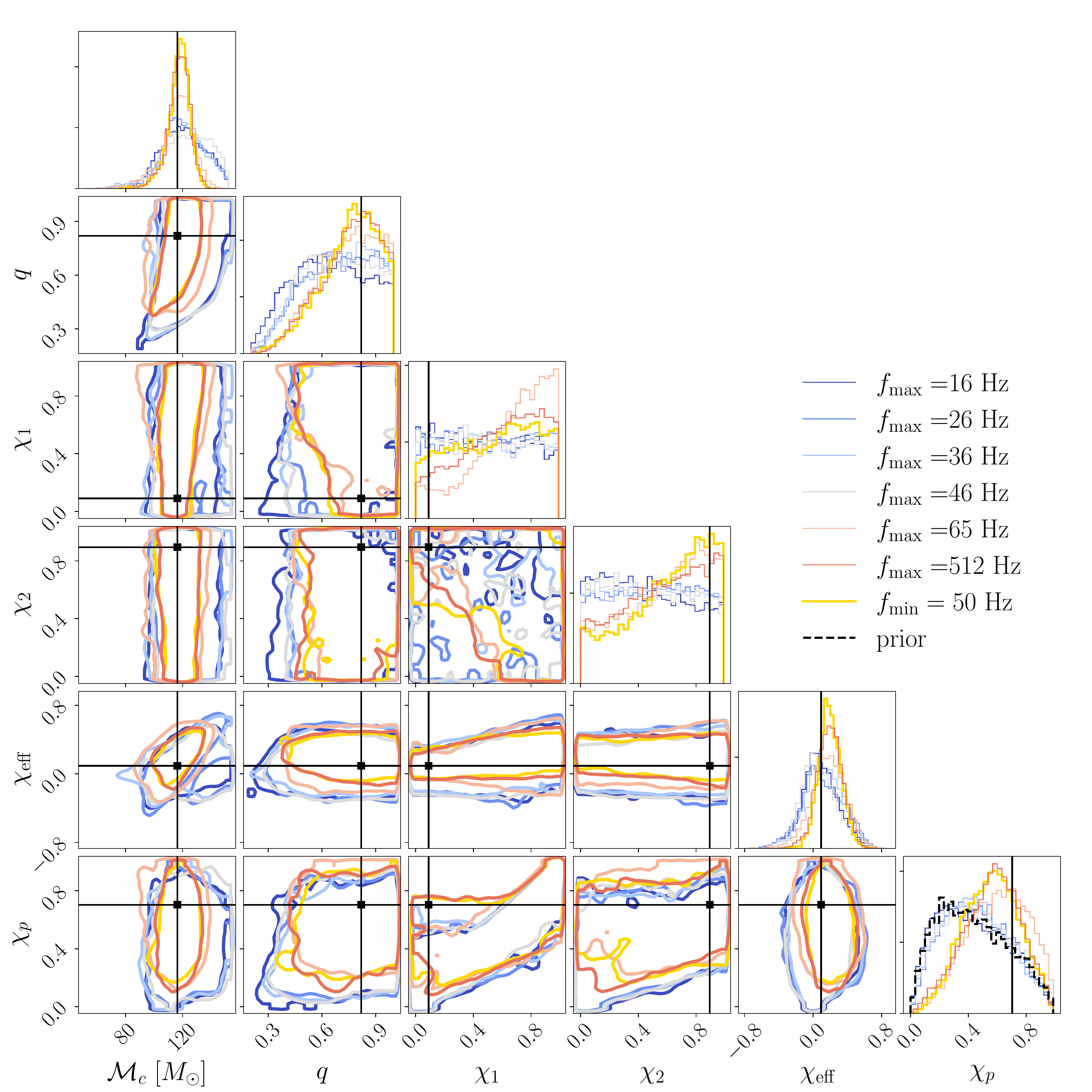}\qquad
\caption{Comparison corner plot for the posteriors obtained for the chirp mass,
mass ratio, component spin magnitudes, and effective aligned and precessing
spin parameters for the GW190521 maximum-likelihood simulation using different
maximum and minimum frequencies.
}
\label{fig:fhigh_corner}
\end{figure*}

The posterior probability distributions obtained using each of the different
maximum frequencies are shown in Fig.~\ref{fig:fhigh_corner}. The posteriors
for all the parameters shown remain largely unchanged as the maximum frequency
is increased until an $f_{\max}$ of $65~\mathrm{Hz}$ is reached. This can be explained due to the
fact that the SNR of each simulated system is kept constant. Even though more inspiral
cycles are added as we increase the analyzed bandwidth, each cycle becomes less
informative since the amplitude of the signal decreases to keep the overall SNR the
same. 

Even for the choices of $f_{\max}=46, 50~\mathrm{Hz}$, which include
frequency content after the merger, the posteriors are qualitatively similar to
those obtained when analyzing only the inspiral frequency content. However, the
posterior constraints improve dramatically for all parameters when $f_{\max}=65~\mathrm{Hz}$ is used, and become further constrained when the full
frequency band is analyzed. Allowing for frequency content up to $65~\mathrm{Hz}$ includes significantly more of the postmerger waveform than either 45 or 50 Hz. This indicates that the ringdown part of the signal,
where the frequency as a function of time changes concavity in the top panel of Fig.~\ref{fig:waveform}, 
 contributes the
most to the measurement of both the mass and spin parameters for such heavy BBH
systems. %

As a verification of the contribution of the postmerger part of the signal to the ability to constrain the spin parameters, we repeat the analysis of the system above but instead of changing the maximum frequency, we limit the minimum frequency to $50~\mathrm{Hz}$. This ensures that the inspiral has been completely removed from the analyzed signal, and any information on the spins must come entirely from the merger and ringdown. The posteriors for the spin parameters obtained with $f_{\min} = 50~\mathrm{Hz}$ are shown in gold in Fig.~\ref{fig:fhigh_corner}. These posteriors are nearly identical to those obtained when analyzing the full frequency band. This demonstrates that the strong dependence of the measurability of the spin parameters on the postmerger part of the signal is not artificially due to the limited sensitivity of the detectors at low frequencies. Rather, the physical richness of the merger and ringdown alone is sufficient to provide a spin constraint equal to that obtained when also including the inspiral in the analysis~\cite{Hughes:2019zmt, Lim:2019xrb}.

\section{Accuracy and precision of spin measurements for high-mass systems}
\label{sec:bias_vs_mass}
In the previous section we have shown that certain spin configurations can lead to informative measurements driven by the postmerger part of the signal for heavy BBH systems. Next, we seek to extend this analysis to generic systems to determine how the measurability of the spin parameters depends on the total mass of the binary. We investigate the effect of systematically varying the total mass for different choices of primary spin tilt, mass ratio, and inclination angle. We consider systems with two different inclination angles, $\theta_{JN}=0.5, 1.4$ (which we refer to as nearly face-on and nearly edge-on, respectively) and two mass ratios, $q = 1, 1/4$. We also use three different primary tilt angles, $\theta_{1/A}=0, \pi/3, \pi/2~\mathrm{rad}$, corresponding to systems with aligned, moderate, and in-plane tilt angles (see the first set of simulations presented in Table~\ref{tab:injections}). 

\begin{table*}[ht!]
\caption{True parameter values for the four different sets of simulations performed for Sections~\ref{sec:bias_vs_mass} and \ref{sec:phases} of this work. In the text, we refer to the two different choices of mass ratio as equal and unequal, and to the two different choices of inclination angle for Simulation 1 as nearly face-on and nearly edge-on.}
\label{tab:injections}
\begin{ruledtabular}
\begin{tabular}{p{1cm} p{1.5cm} p{2cm} p{1.5cm} p{1.5cm} p{1.5cm} p{1.2cm} p{1.2cm} p{2cm} p{2cm} }
    Set & SNR & $M_{\mathrm{tot}}~[M_{\odot}]$ & $q$ & $\chi_{1}$ & $\theta_{1}~\mathrm{[rad]}$ & $\chi_{2}$ & $\theta_{2}~\mathrm{[rad]}$ & $\phi_{12}~\mathrm{[rad]}$ & $\theta_{JN}~\mathrm{[rad]}$  \\
    \midrule
    1 & 30& $75 - 250$ & 1, 0.25 & 0.8 & $0, \pi/3, \pi/2$ & 0 & -- & -- & 0.5, 1.4 \\
    2 & 30& $75 - 250$ & 1, 0.25 & 0.8 & $\pi/3$ & 0.2 & $\pi/6$ & $0, \pi/2, \pi$ & 0.5\\
    3 & 30& $75 - 300$ & 1, 0.25 & 0.8 & $\pi/3$ & 0.7 & $\pi/6$ & $0, \pi/2, \pi$ & 0.5\\
    4 & 60& $75 - 300$ & 1, 0.25 & 0.8 & $\pi/3$ & 0.2 & $\pi/6$ & $0, \pi/2, \pi$ & 0.5\\
\end{tabular}
\end{ruledtabular}
\end{table*}

In this section, we
sort the binary components by their dimensionless spin magnitudes instead of their masses, as is customary. We use the subscript $A$ ($B$) to indicate the highest (lowest)-spinning black hole~\cite{Biscoveanu:2020are}. Because we include equal-mass systems in our simulations, this sorting helps break the degeneracy between the spins of the two black holes. In the unequal-mass case, the heavier black hole is also the one with higher spin, so there is no distinction between the two sorting methods. The true values for the remaining parameters which are kept the same across all the simulations presented in this section are listed in Table~\ref{tab:same_params}. 

While the simulated systems presented in this section are not necessarily tuned to the optimal configuration found in the previous section, the accuracy and precision of the spin measurements are comparable to those previously presented. The conclusions we draw in this section---particularly with respect to comparisons between different inclination and primary tilt angles---are thus robust across a broad range of systems. 

\begin{table}
\caption{True values for the parameters common to all four simulation sets described in Sections~\ref{sec:bias_vs_mass} and \ref{sec:phases}.}
\label{tab:same_params}
\begin{ruledtabular}
\begin{tabular}{lll}
    Parameter & Symbol & Value\\
    \midrule
    Azimuthal precession cone angle & $\phi_{JL}$ & 6.28 \\
    Polarization angle & $\psi$  & 0.28  \\
    Coalescence phase & $\phi_{c}$  & 4.70 \\
    Right ascension & $\alpha$ & 1.09 \\
    Declination & $\delta$ & 1.47 \\
    Coalescence GPS time & $t_{c}$ & 1126259642 s \\
\end{tabular}
\end{ruledtabular}
\end{table}

\subsection{Equal-mass systems}
\begin{figure*}
	\centering
	\includegraphics[width=0.93\textwidth]{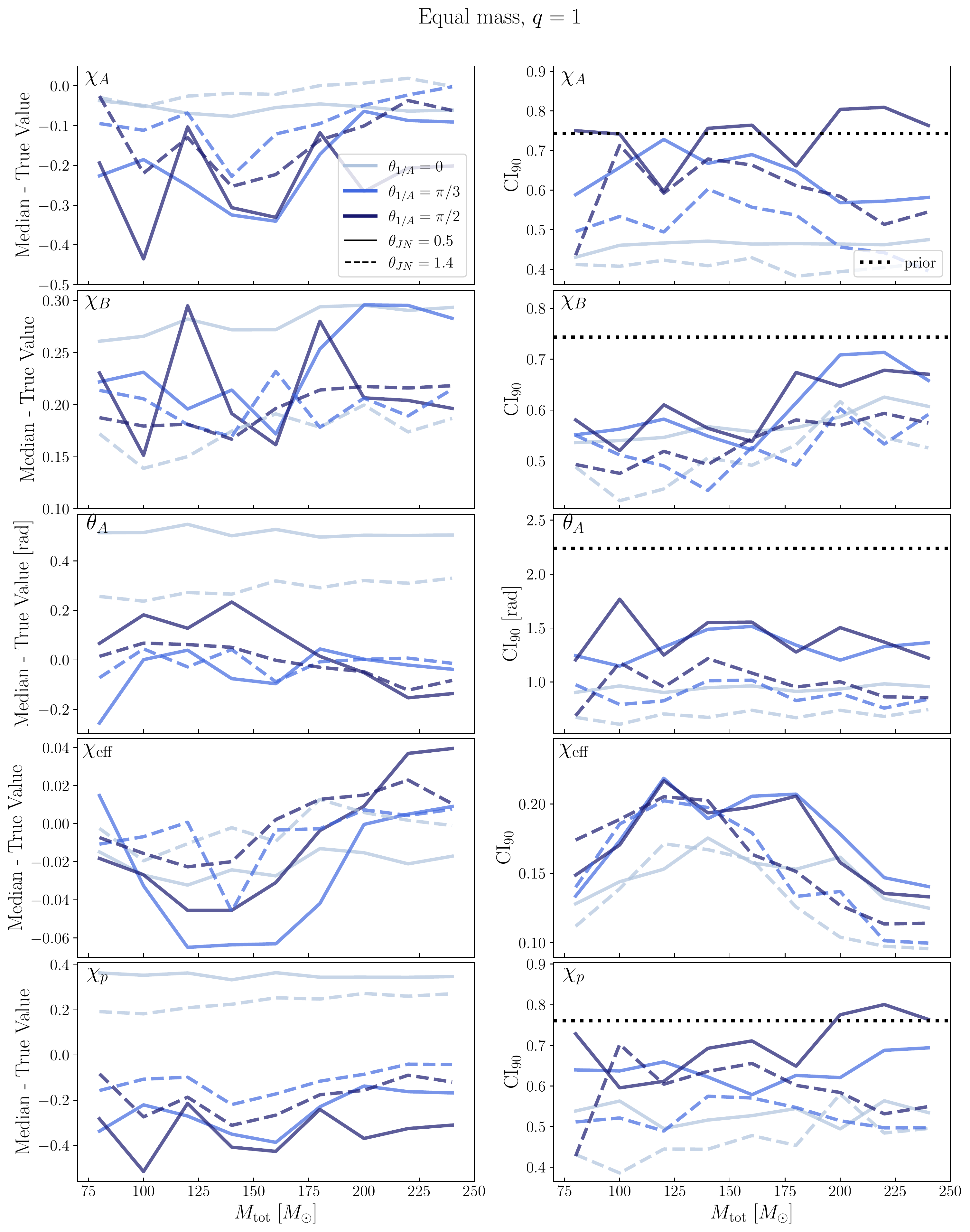}\qquad
	\caption{The bias and width of the spin-sorted spin magnitudes, tilt of the highest-spinning black hole, and effective aligned and precessing spins as a function of the total mass for systems with equal mass. The bias is defined as the difference between the median of the posterior and the true value, while the width is the symmetric 90\% credible interval. The secondary black hole is nonspinning, and the remaining true parameter values are given in Tables~\ref{tab:injections} and \ref{tab:same_params}. Aligned spin systems with $\theta_{1/A}=0$ are shown in light blue, those with $\theta_{1/A}=\pi/3$ in medium-blue, and those with in-plane primary tilt, $\theta_{1/A}=\pi/2$, in dark blue. Systems observed nearly face-on with $\theta_{JN}=0.5~\mathrm{rad}$ are represented with a solid line, and those observed nearly edge-on with $\theta_{JN}=1.4~\mathrm{rad}$ are represented with a dashed line. The dotted black lines show the width of the 90\% CI for the prior for all parameters except \chieff, for which the prior width of 0.84 is much larger than the posterior widths.}
	\label{fig:equal_mass}
\end{figure*}

Figure~\ref{fig:equal_mass} shows the bias and uncertainty in the recovered component spin magnitudes, $\chi_{A}$ and $\chi_{B}$, tilt angle of the highest-spinning black hole, $\theta_{A}$, and effective aligned and precessing spins, \chieff and $\chi_{p}$, as a function of the true total mass of the system for all simulated systems with $q=1$. We define the bias as the difference between the median of the posterior and the true value, and the uncertainty is represented by the width of the 90\% credible interval (CI) of the posterior. Thus, the bias is a measure of the accuracy of the posterior, and the 90\% CI width is a measure of the precision. The bias is ill-defined for $\theta_{B}$ since this quantity is undefined for a nonspinning component; consequently, this parameter is unconstrained, and its posteriors just return the prior for all simulations. 

For equal-mass spin-aligned systems ($\theta_{1/A}=0$; light blue lines in Fig.~\ref{fig:equal_mass}), there is no significant trend in the accuracy or precision of the measurement of the component spin magnitudes or tilts with the mass of the system for either of the two inclination angles we simulated.  The large positive bias in the $\theta_{A}$ posterior for these systems is due to the fact that the measurement of $\theta_{A}$ is not very informative and is hence driven by the uniform-in-$\cos{\theta_{A}}$ prior, which peaks at $\theta_{A}=\pi/2$, so the median is always higher than the true value of $\theta_{A}=0$.
The posteriors for both $\chi_{A}$ and $\theta_{A}$ are more constrained and less biased for systems with nearly edge-on inclinations (dashed lines in Fig.~\ref{fig:equal_mass}), however. This is a generic feature of all our simulations, and can be explained due to the fact that, for a fixed SNR, viewing the system nearly edge-on enhances the effect of precession on the observed waveform. In some configurations where $\cos{\theta_{JL}} < \sin{\theta_{JN}}$ (see the right panel of Fig.~\ref{fig:schematic}), the observer can see both sides of the orbital plane as it oscillates~\cite{Vitale:2014mka}.

For systems with moderate primary spin tilt, $\theta_{1/A}=\pi/3$, shown in the medium-blue lines in Fig.~\ref{fig:equal_mass}, the constraints on $\chi_{A}$ improve slightly at higher total masses, although the same trend is not observed in $\theta_{A}$ for this choice of tilt angle. 
The width of the posterior is largest at about $M_{\mathrm{tot}}\approx 160~M_{\odot}$, corresponding to a local minimum in the bias as a function of mass for $\chi_{A}$. This trend is observed for both inclination angles, although the posteriors are again more constrained for the nearly edge-on simulations. For systems with in-plane primary spin, this trend is only observed at nearly edge-on inclination angles (dark blue dashed line), while the width of the 90\% credible interval for the nearly face-on systems (dark blue solid line) remains largely constant with mass. 

The posteriors of the lowest-spinning object do not exhibit a turn-over in the width of the 90\% CI for either $\theta_{1/A}=\pi/3$ or $\theta_{1/A}=\pi/2$. Instead, the width increases slightly as a function of mass. The bias for $\chi_{B}$ is roughly constant as a function of both mass and primary tilt angle. On the other hand, both the accuracy and the precision of the $\chi_{A}$ posteriors improve as the primary spin tilt decreases (as the line color gets lighter in the top panel of Fig.~\ref{fig:equal_mass}). This is indicative of the fact that that it is easier to measure the component spins for aligned-spin systems, where the absence of precession in the waveform is informative, as seen in Ref.~\cite{Vitale:2016avz}, since only the well-measured aligned-spin component, $\chi_{A, z}$, contributes.

\begin{figure}
	\centering
	\includegraphics[width=0.5\textwidth]{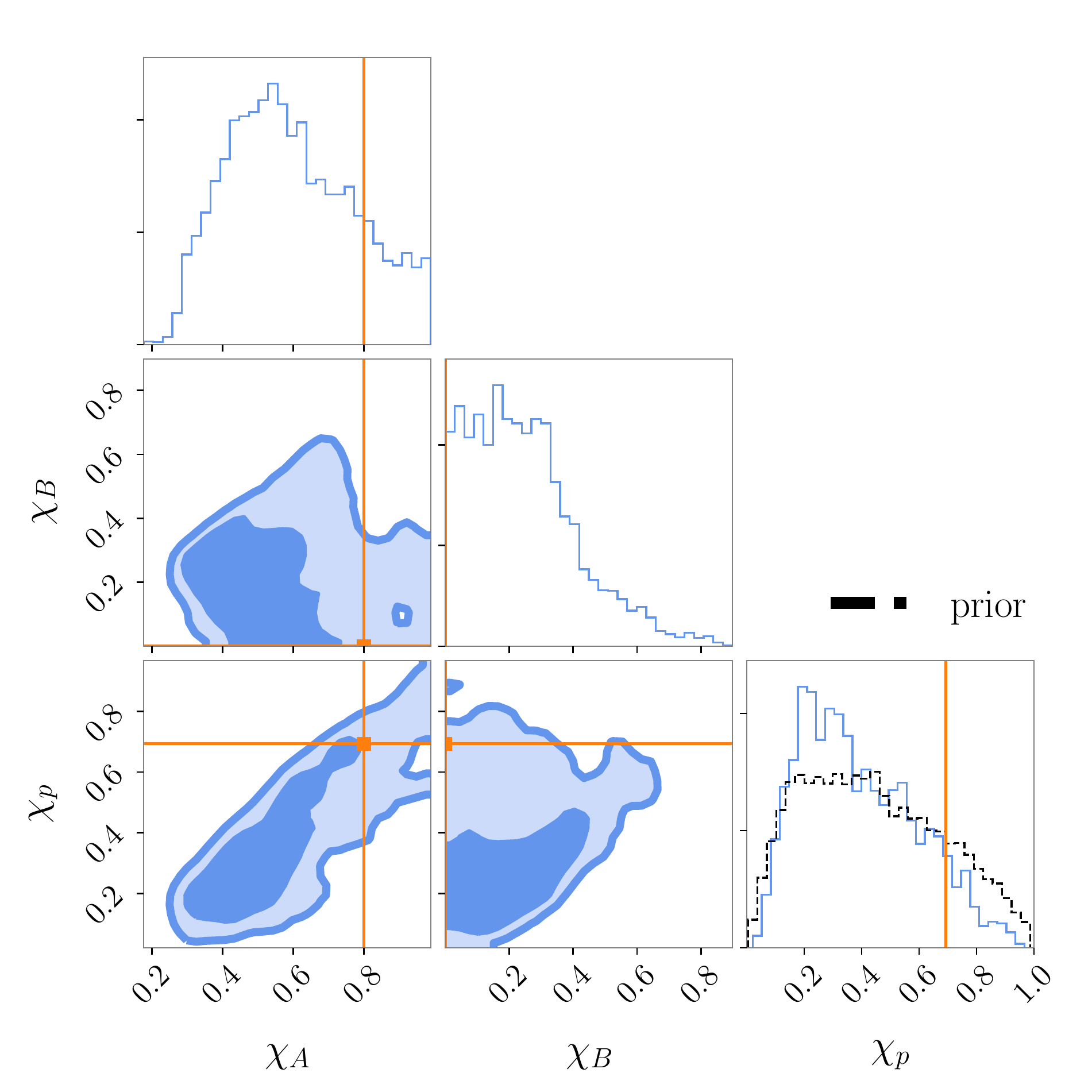}\qquad
	\caption{Corner plot showing the $\chi_{A}$, $\chi_{B}$, and $\chi_{p}$ posteriors for an equal-mass system with nonzero primary tilt. The light shaded region shows the 90\% CI, and the darker region shows the 50\% CI. The prior for $\chi_{p}$ is shown in the black dashed line, and the true parameter values are given by the vertical orange lines.}
	\label{fig:chiA_bias_corner}
\end{figure}

Comparing the scale of the first and second panels on the right side of Fig.~\ref{fig:equal_mass} shows that $\chi_{B}$ is better constrained than $\chi_{A}$ for systems with nonzero primary tilt. The inference for these sources consistently recovers lower values for $\chi_{A}$ than the true value, as indicated by the negative values of the bias. This is due to a degeneracy with $\chi_{p}$, which is shown in the corner plot in Fig.~\ref{fig:chiA_bias_corner}. The low-$\chi_{p}$ tail driven by the prior (black dashed line) pulls the spin magnitude to lower values. This preference for low spins translates into a better constraint on the spin magnitude of the lowest-spinning black hole, since high spin values can be entirely ruled out for that object but not for the highest-spinning object. 

This bias in the magnitude of $\chi_{A}$ towards low values is also present in the posteriors for \chieff and $\chi_{p}$, as shown in the lower two panels on the left side of Fig.~\ref{fig:equal_mass}. The width of the 90\% CI for \chieff decreases at the highest masses regardless of primary tilt or inclination angle. We observe no significant trend in the width of the posteriors as a function of mass for $\chi_{p}$. The large positive bias in $\chi_{p}$ for aligned-spin systems follows from the corresponding bias in $\theta_{A}$, and is due to the fact that the true value of $\chi_{p}=0$ for those systems, so the median is always larger since the prior for $\chi_{p}$ is peaked away from $\chi_{p}=0$. The posteriors for the systems observed nearly edge-on are less biased and better constrained for all choices of primary tilt compared to those for systems observed nearly face-on for both \chieff and $\chi_{p}$. We emphasize that the biases we comment on above are driven by the effects of the prior shape and projecting the posteriors into one dimension to calculate the median. The maximum likelihood point in the full 15-dimensional parameter space should coincide with the true parameter values since we are not adding Gaussian noise to the signal. Hence, any bias in the posterior median away from the true value occurs when the impact of the prior exceeds that of the likelihood.

\subsection{Unequal-mass systems}
\begin{figure*}
	\centering
	\includegraphics[width=0.93\textwidth]{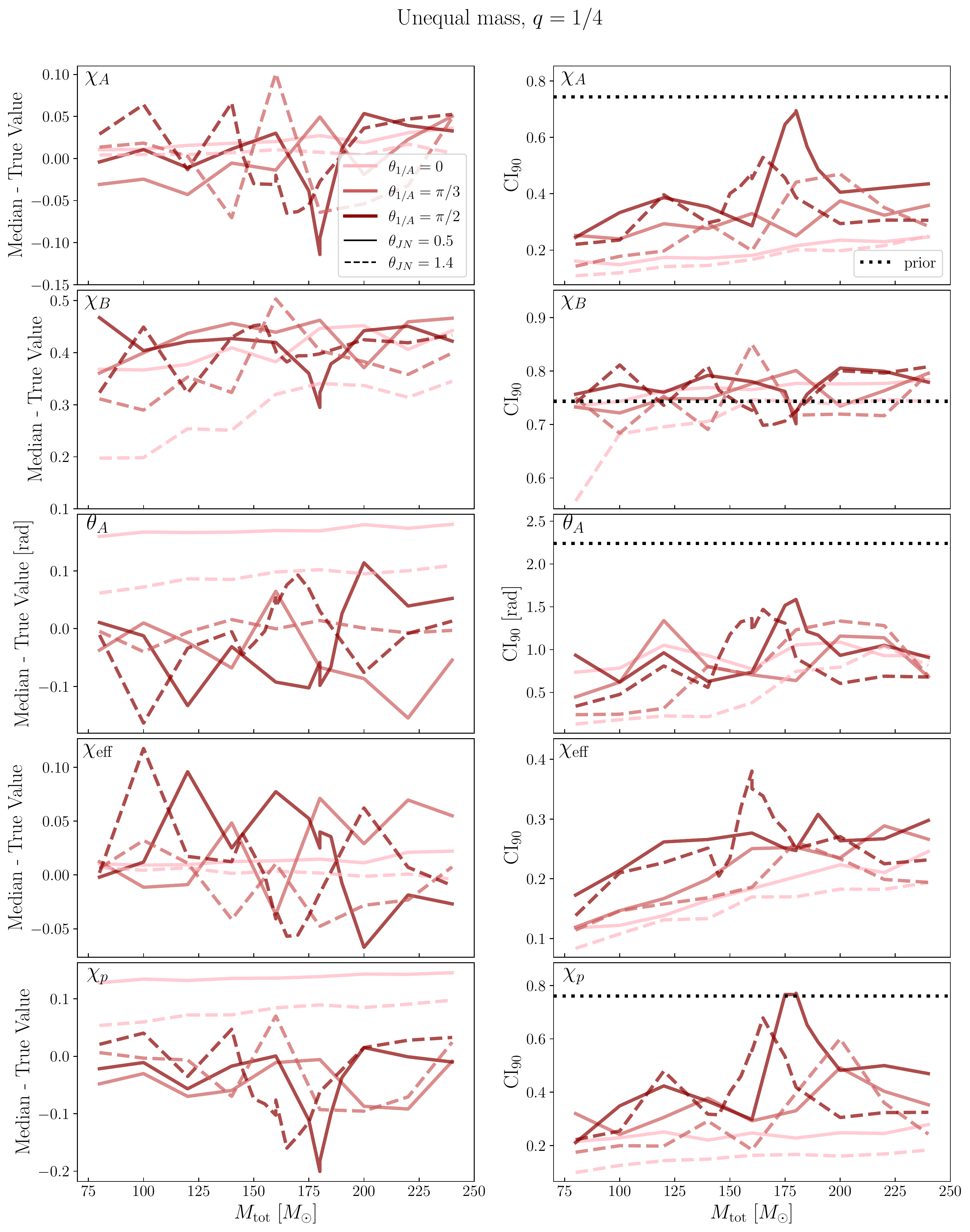}\qquad
	\caption{The bias and width of the spin-sorted spin magnitudes, tilt of the highest-spinning black hole, and effective aligned and precessing spins as a function of the total mass for systems with unequal mass, $q=1/4$. The bias is defined as the difference between the median of the posterior and the true value, while the width is the symmetric 90\% credible interval. The secondary black hole is nonspinning, and the remaining true parameter values are given in Tables~\ref{tab:injections} and \ref{tab:same_params}. Aligned spin systems with $\theta_{1/A}=0$ are shown in light red, those with $\theta_{1/A}=\pi/3$ in medium-red, and those with in-plane primary tilt, $\theta_{1/A}=\pi/2$ in dark red. Systems observed nearly face-on with $\theta_{JN}=0.5~\mathrm{rad}$ are represented with a solid line, and those observed nearly edge-on with $\theta_{JN}=1.4~\mathrm{rad}$ are represented with a dashed line. The dotted black lines show the width of the 90\% CI for the prior for all parameters except \chieff, for which the prior width of 0.84 is much larger than the posterior widths.}
	\label{fig:unequal_mass}
\end{figure*}

The bias and uncertainty for the same spin parameters presented above are shown for our $q=1/4$ simulations in Fig.~\ref{fig:unequal_mass}. The most salient features of our analysis of unequal-mass systems are the apparent outliers in the width of the 90\% credible interval of $\chi_{A}$ and $\theta_{A}$ at ${\sim}180~M_{\odot}$ for systems with in-plane primary tilt observed nearly face-on (dark red solid line) and at ${\sim}160~M_{\odot}$ for systems with in-plane primary tilt observed nearly edge-on (dark red dashed line). We verify that these apparent outliers are not due to sampling issues by repeating the inference with a different random seed for the stochastic sampler and by performing more simulations with a finer mass spacing ($5~M_{\odot}$) for the region with width $40~M_{\odot}$ centered on the outlier value in each case. These additional simulations, which are included among the systems plotted in Fig.~\ref{fig:unequal_mass}, demonstrate that the outliers represent local maxima in the width of the 90\% CI as a function of mass, with values $\gtrsim 0.2$ in excess of those for the lowest and highest masses. As we will describe in  Section~\ref{sec:outliers}, these apparent outliers are a consequence of the choice to define the true spin parameter values at a fixed reference frequency of $25~\mathrm{Hz}$ for all systems regardless of total mass. This results in the compared systems having different values of the spin angles at the same point in their evolution as the total mass increases. We emphasize, however, that the presence of these apparent outliers does not affect the conclusions drawn when comparing the results across different values of inclination, primary tilt, and mass ratio, which are discussed in the rest of this section.

Both the accuracy and the precision of the measurements of the spin components for the highest-spinning object improve considerably for unequal-mass systems compared to those with equal mass. This is due to the fact that the unequal mass serves to break parameter degeneracies in the system~\cite{Kidder:1995zr}. The constraints for the lowest-spinning object conversely suffer due to this degeneracy breaking, as all the spin information obtained pertains to the highest-spinning object in this case, and the secondary spin measurements are completely uninformative.

The unequal-mass aligned-spin systems, represented by the light red lines in Fig.~\ref{fig:unequal_mass}, show a gradual broadening of the posteriors on the magnitude and tilt of the highest-spinning black hole as a function of mass. A similar trend is observed for those systems with moderate primary spin tilt, shown in medium-red. While the results presented in Section~\ref{sec:fmax} suggest that the spin information is dominated by the merger and ringdown parts of the signal for heavy BBH, it is possible that having access to a longer inspiral still leads to improvements in the measurability of spin relative to a system with a shorter inspiral. Thus, the degradation of the component spin measurements for the highest-spinning black hole as a function of increasing total mass could be explained due to the presence of fewer inspiral cycles in the band of the gravitational-wave detector as the mass increases. That is, as the total mass increases, even though the spin information gained from the merger-ringdown becomes relatively more significant, the absolute spin information gain still decreases because fewer cycles fall in the detector band. The same explanation would apply for the increase in the width of the 90\% CI for $\chi_{B}$ as a function of mass for equal-mass systems. We leave detailed investigation of this hypothesis to future work.

The bias in the \chieff posteriors for unequal-mass systems varies significantly as a function of mass, although the scale of the bias is smaller than that of the component spin magnitudes. $\chi_{p}$ is much less biased for unequal compared to equal masses, indicating that it is easier to accurately capture precession via $\chi_{p}$ when the parameter degeneracies in the system are broken. This is due to the fact that there are post-Newtonian terms proportional to the difference in component masses that disappear for equal-mass systems, making the spin parameters harder to measure in this case~\cite{Kidder:1995zr}.

The \chieff posteriors for unequal-mass systems do not show the same decrease in the width of the 90\% credible interval at the highest masses as the equal-mass systems, with the exception of the source with in-plane primary spin observed edge-on that exhibits the outlier in $\chi_{A}$ (dark red dashed line). The same outlier and trend is visible in \chieff for this system, but not for the source observed nearly face-on (dark red solid line in the penultimate panel on the right side of Fig.~\ref{fig:unequal_mass}). For the nearly face-on source, the increased uncertainties in the parameters going into the numerator and denominator of Eq.~\eqref{eq:chi_eff} cancel out, since the mass ratio suffers from a similar increase in the width of the 90\% credible interval (see Fig.~\ref{fig:mass_bias} in Appendix~\ref{ap:masses}). The outliers in the systems with in-plane primary spin observed at both inclination angles are present in the width of the 90\% credible interval for $\chi_{p}$, however. There is another smaller outlier in $\chi_{p}$ at around $M_{\mathrm{tot}}=180~M_{\odot}$ for the unequal-mass systems with moderate primary tilt observed nearly edge-on (medium-red dashed line), that can also be seen in the widths of the posteriors for $\chi_{A}$ and $\theta_{A}$. With the exception of these outliers, the width of the 90\% credible interval for both the \chieff and $\chi_{p}$ posteriors gradually increases with the total mass of the system.

\subsection{Outliers}
\label{sec:outliers}
The apparent $q = 1/4$ outliers mentioned in the previous section show a deterioration in the precision of the measurements of $\chi_{A}$, \chieff, and $\chi_{p}$ at masses in the range of $\sim 160-180~M_{\odot}$, and then a subsequent improvement at the highest masses. This goes against the naive expectation that the spin measurements should get worse for higher total masses since fewer inspiral cycles are accessible in the band of the interferometers as the total mass of the system increases. The constraints on the mass parameters for these systems, shown in Fig.~\ref{fig:mass_bias} in Appendix~\ref{ap:masses}, exhibit a similar trend. 

The local maxima in the widths of the $\chi_{A}$ posteriors correspond to local minima in the widths of the $\chi_{B}$ posteriors, indicating that one is measured better at the expense of the other, although the peaks are still visible in the posteriors for the effective spins since the scale of the upward deviation is much larger than that of the downward deviation. These features are therefore not an artifact of the spin sorting, as the definitions of \chieff and $\chi_{p}$ do not change between sortings. The peaks in the widths as a function of mass also correspond to minima in the bias for $\chi_{A}$ and $\chi_{p}$, indicating that for these systems the inference recovers spins that are systematically lower than the true values and not well constrained. We note that the two sets of simulations that exhibit this outlier behavior correspond to sources with the same intrinsic parameters just viewed at different inclination angles.

\begin{figure}
\centering
\includegraphics[width=0.5\textwidth]{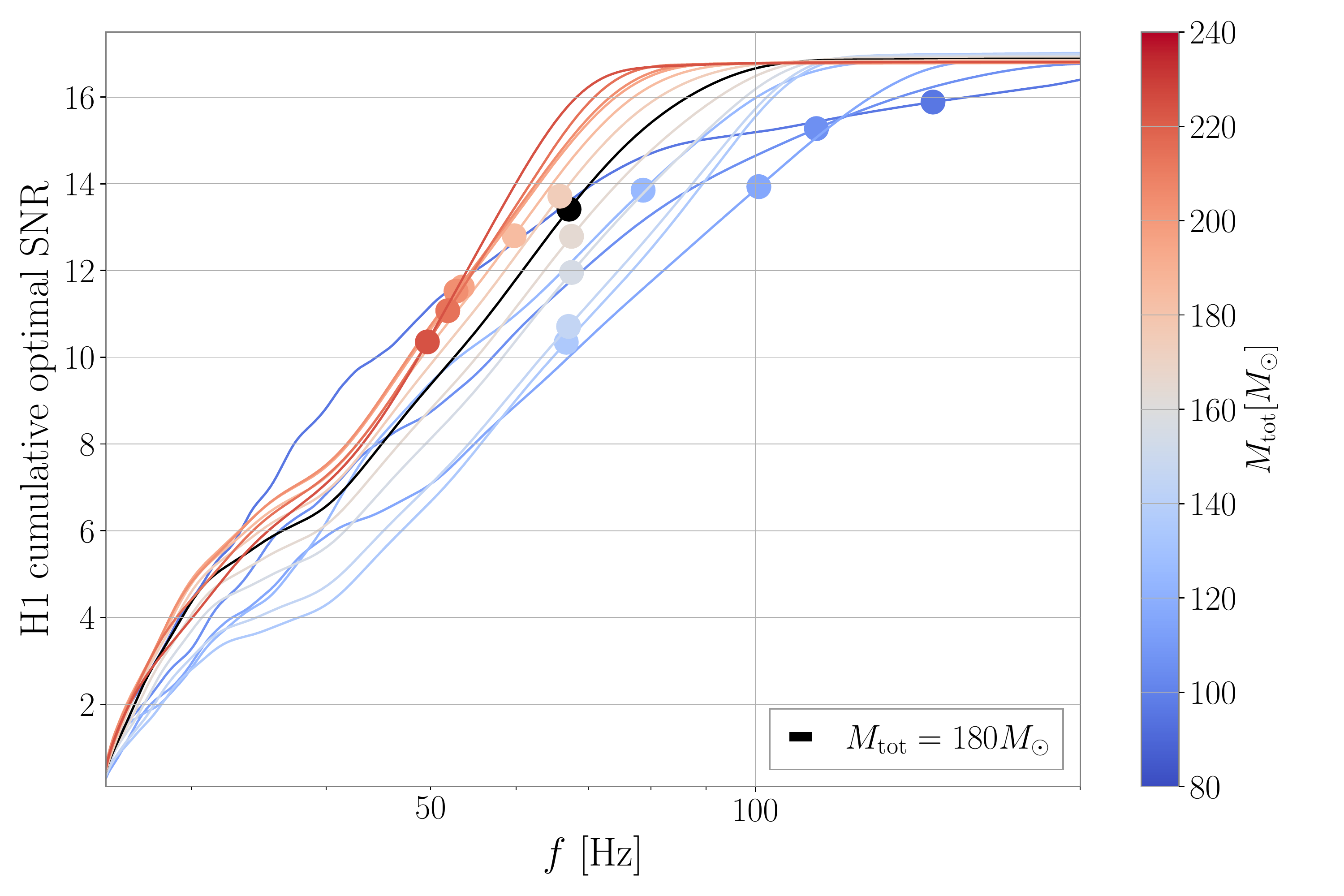}
\includegraphics[width=0.5\textwidth]{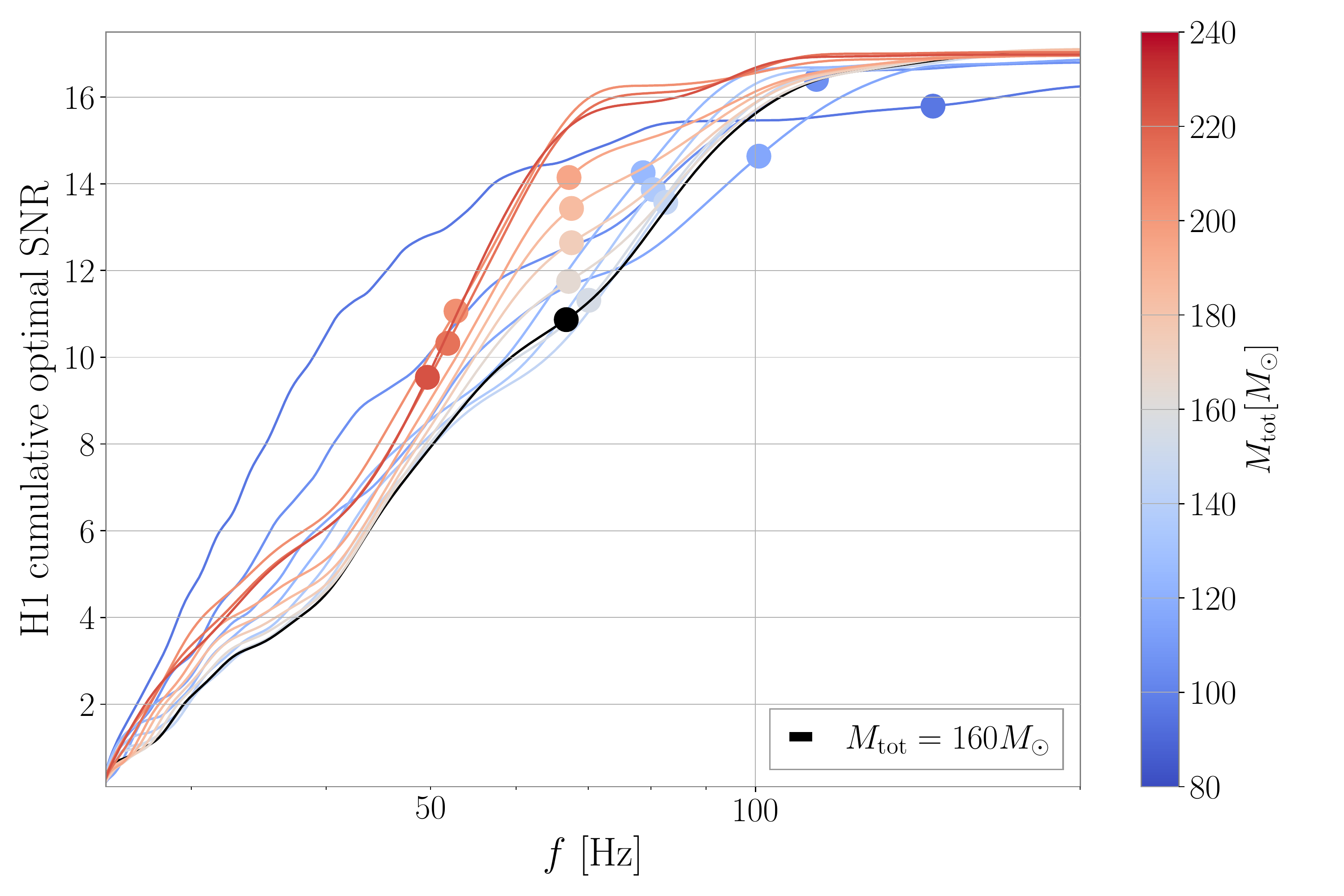}
\caption{Cumulative SNR as a function of frequency in the Hanford
interferometer for different total masses for the simulations observed nearly
face-on (top) and nearly edge-on (bottom). The marker shows the merger
frequency for each total mass, and the outlier mass is shown in black.
}
\label{fig:cumulative_snr}
\end{figure}

In Fig.~\ref{fig:cumulative_snr}, we plot the cumulative SNR in the Hanford interferometer as a function of frequency for the different total masses we simulated. For the nearly face-on case (top panel), the apparent outlier at $180~M_{\odot}$ occurs when the inspiral contributes most to the total SNR of the signal (with the exception of the lowest mass sources). However, looking instead at the nearly edge-on case, the opposite is true; the outlier at $160~M_{\odot}$ occurs when the merger contributes most to the total SNR of the signal (except for the highest-mass sources). This feature is consistent across the SNR distributions for all three interferometers, and indicates that the apparent outliers cannot simply be attributed to one part of the signal contributing most of the spin information for these systems. 

\begin{figure}
	\centering
	\includegraphics[width=0.5\textwidth]{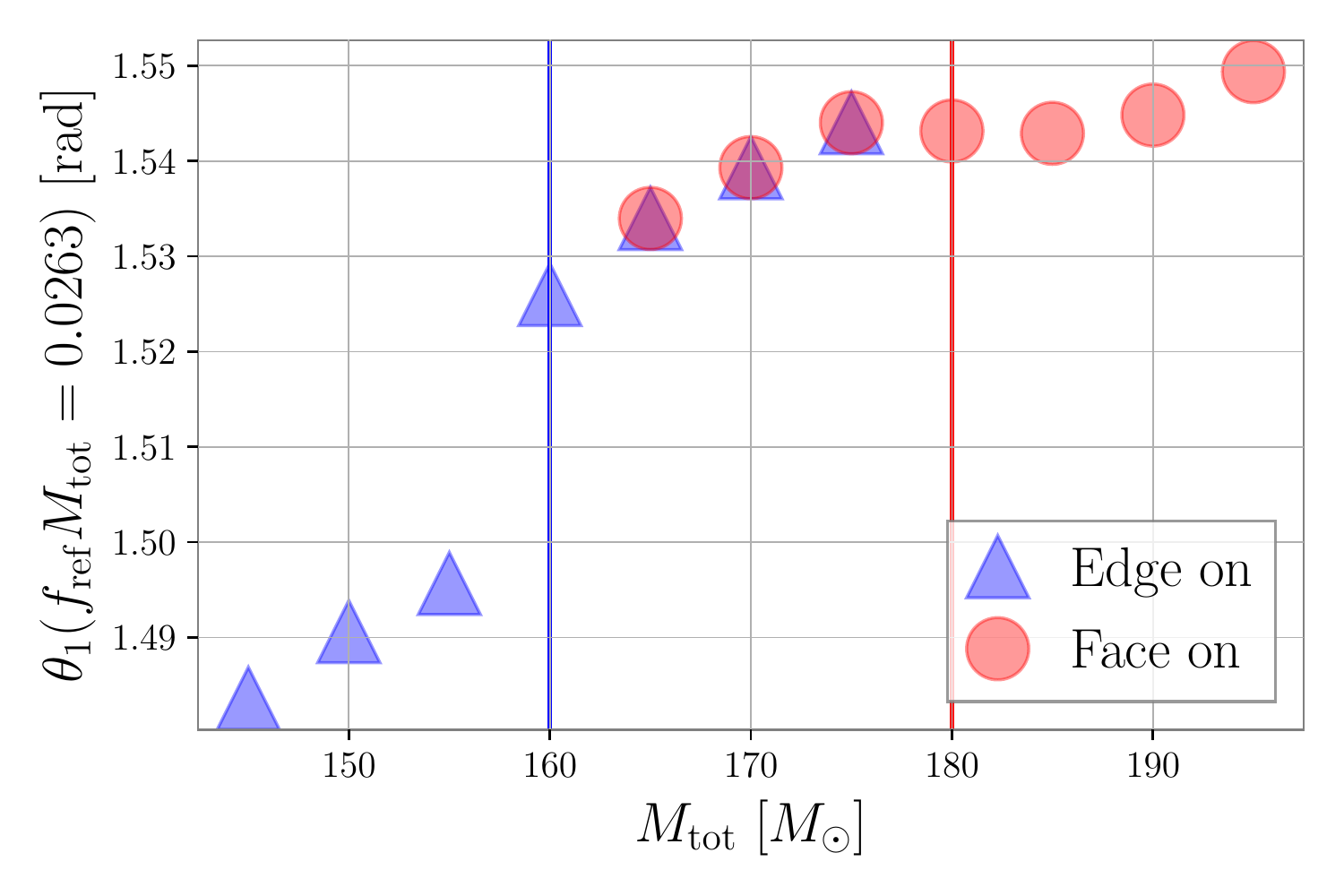}
	\caption{Tilt angle at a fixed dimensionless reference frequency, $f_{\mathrm{ref}}M_{\mathrm{tot}}=0.0263$, as a function of mass for some of the systems with masses in the vicinity of the two outliers. The nearly edge-on systems are shown in blue, and the nearly face-on systems in red. The two overlap for the common masses because they correspond to the same intrinsic system viewed at different angles by the observer. The masses of the outlier systems are indicated with vertical lines.}
	\label{fig:rescaled_tilts}
\end{figure}

Instead, these plots highlight that the merger frequency does not monotonically decrease as a function of total mass for these systems, since the simulations were performed at a fixed reference frequency of 25~Hz. This reference frequency corresponds to different stages in the evolution of the source for systems of different masses, so the primary spin tilt angle will be different at a fixed dimensionless frequency, $f_{\mathrm{ref}}M_{\mathrm{tot}}$, with $M_{\mathrm{tot}}$ in units of seconds. This means the simulated binaries are not the same up to a total mass scaling, but rather correspond to different physical systems with distinct spin parameters. The tilt angles of some of the systems with masses in the vicinity of the two outliers are shown in Fig.~\ref{fig:rescaled_tilts} at $f_{\mathrm{ref}}M_{\mathrm{tot}}=0.0263$, approximately one cycle before merger. That the two orientations have overlapping tilt angles in the common mass range is expected since they represent the same intrinsic system viewed at different inclination angles. While the overall scale of the variation in the primary tilt angle is less than a tenth of a radian, the apparent outliers (indicated with vertical lines) occur at inflection points in the tilt as a function of mass. This suggests that the exact spin configuration can have a significant effect on the measurability of the spin parameters for high-mass systems, reinforcing the finding from Section~\ref{sec:high_chip}.

\begin{figure*}
	\centering
	\includegraphics[width=\textwidth]{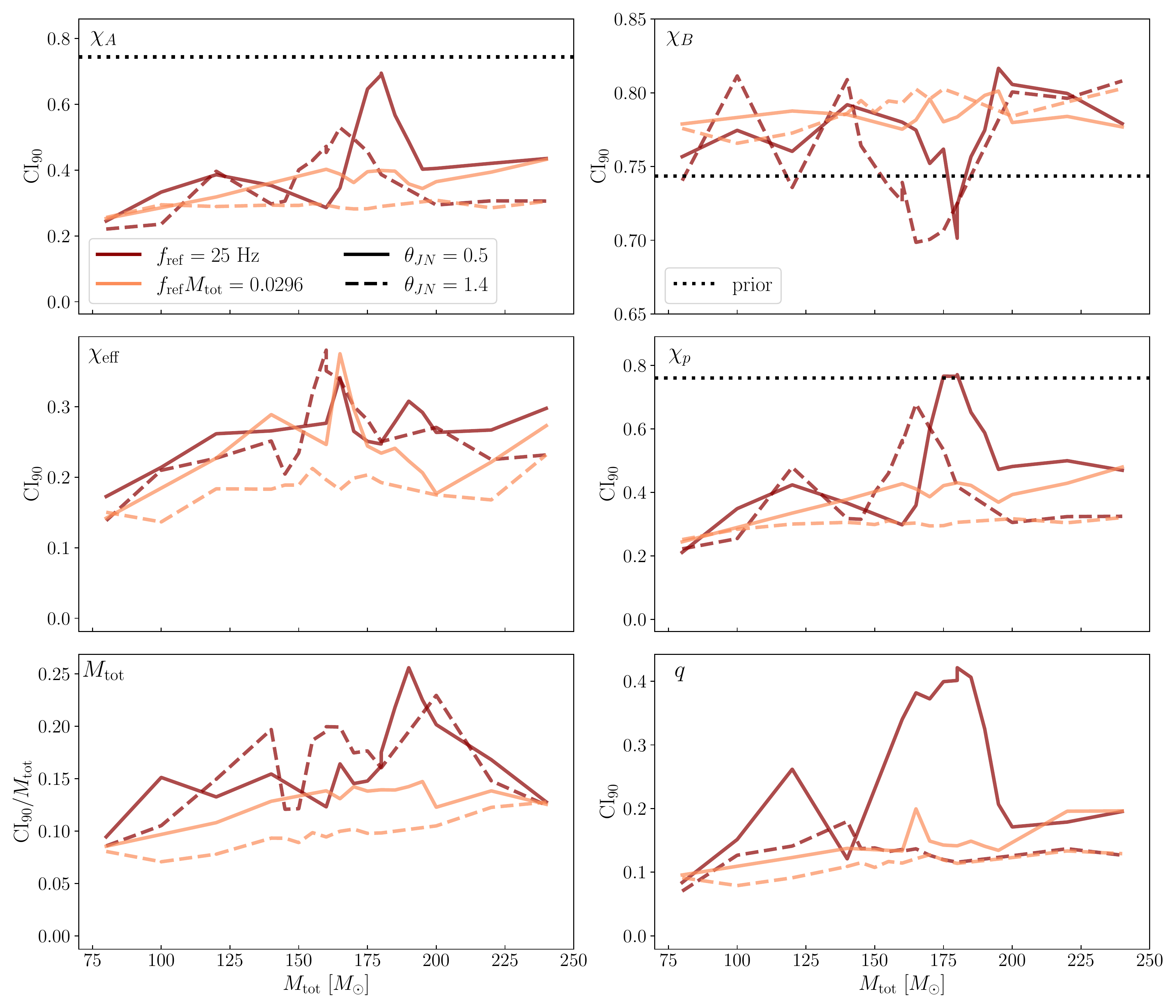}
	\caption{Width of the 90\% credible interval as a function of total mass for the original systems with outliers at fixed $f_{\mathrm{ref}}=25~\mathrm{Hz}$ shown in dark red compared to the simulations with the same parameters but at fixed dimensionless reference frequency $f_{\mathrm{ref}}M_{\mathrm{tot}}=0.0296$ in orange. The systems viewed nearly edge-on are indicated with dashed lines and those observed nearly face-on with solid lines. For the total mass, the credible interval width is normalized by the value of the total mass of each system so that this quantity is dimensionless.}
	\label{fig:fixed_frefMtot}
\end{figure*}

To verify that the apparent outliers are due to the choice of fixed reference frequency, we repeat the simulations for the configurations with outliers---namely, unequal-mass systems with in-plane primary tilt observed both nearly edge-on and nearly face-on---but, instead of using a fixed reference frequency of $25~\mathrm{Hz}$, we use a fixed dimensionless reference frequency of $f_{\mathrm{ref}}M_{\mathrm{tot}}=0.0296$. This value was chosen so that the waveform for the highest-mass system we simulated, $M_{\mathrm{tot}}=240~M_{\odot}$, could be generated with $f_{\mathrm{ref}}=f_{\min}=25~\mathrm{Hz}$. The widths of the 90\% credible intervals for the component spin magnitudes, effective spins, and masses are shown in Fig.~\ref{fig:fixed_frefMtot} for both the original fixed reference frequency (shown in dark red) and the new fixed dimensionless reference frequency (shown in orange). While the trend in the posterior width as a function of mass is not completely smoothed out when using the fixed dimensionless reference frequency, the outliers are no longer present. Instead, the width increases nearly monotonically as a function of total mass for all parameters except $\chi_{B}$. For this parameter, the width is instead roughly constant when comparing the scale of the variation to that of the other parameters. The posteriors for $\theta_{A}$ with the new fixed dimensionless reference frequency exhibit similar behavior. The large excursions in the bias at the outlier mass for the component spin magnitudes and effective spins in Fig.~\ref{fig:unequal_mass} are also smoothed out. 

By using a fixed dimensionless reference frequency, increasing the total mass of the system does not change the relative phasing of the waveform, but only the overall amplitude and time-frequency scaling. This results in the expected behavior that for a fixed spin configuration, more massive systems have fewer cycles in-band and hence the constraint on the spin parameters gradually worsens as the total mass of the system increases. However, these apparent outliers demonstrate that even small changes in the spin configuration, of the order of a tenth of a radian in the tilt angle, can lead to substantial differences in the posteriors for the spin parameters.

To determine whether these apparent outliers are unique to the NRSur7dq4 waveform, we repeat the parameter estimation for the original systems with a fixed reference frequency of $25~\mathrm{Hz}$ with two additional waveform models, IMRPhenomPv2~\cite{Hannam:2013oca, Husa:2015iqa, Khan:2015jqa} and IMRPhenomXPHM~\cite{Pratten:2020ceb, Pratten:2020fqn, Garcia-Quiros:2020qpx}. Detailed results and comparisons of the three waveform models are presented in Appendix~\ref{sec:waveforms}. In general, we find that these three waveform models yield different results for both the accuracy and precision of the spin and mass parameters.

\section{Measurement of azimuthal spin angles}
\label{sec:phases}

Measurements of the azimuthal spin angles for binary black hole systems can carry important information about their formation channels and evolution~\cite{Gerosa:2013laa, Gerosa:2014kta}. For example, binaries formed in isolation may become locked in a ``spin-orbit resonance'', where the orbital angular momentum and component spin vectors are coplanar and jointly precess around the total angular momentum~\cite{Schnittman:2004vq}. In this configuration, the projections of the two component spins onto the orbital plane are either aligned or antialigned, so that the angle between them $\phi_{12}=0, \pm \pi$. For stellar binaries with significant supernova natal kicks and stellar tides, an increasing fraction of systems will get captured into this resonant configuration as the binary orbital separation decreases, where $\phi_{12}$ librates about $0$ or $\pm\pi$ instead of circulating over the full allowable range of values as the binary evolves. Whether the in-plane spin components are aligned or antialigned depends on the efficiency of mass transfer between the first-born black hole and its remaining stellar companion~\cite{Gerosa:2013laa}.

The detectability of such spin-orbit resonances has previously been studied in \cite{Gupta:2013mea, Gerosa:2014kta, Trifiro:2015zda, Afle:2018slw}. While Refs.~\cite{Gupta:2013mea, Gerosa:2014kta, Afle:2018slw} computed the overlap between the waveforms of different systems to determine if the two resonant configurations are detectable and distinguishable from each other and from non-resonant configurations, Ref.~\cite{Trifiro:2015zda} conducted full parameter estimation on low-mass resonant BBH sources using the SpinTaylorT4 model, which includes the effects of spin precession but only in the inspiral regime relying on the post-Newtonian approximation~\cite{Buonanno:2002fy, Buonanno:2009zt}.
Here we investigate the measurability of $\phi_{12}$ defined at 25 Hz for massive BBH systems using the NRSur7dq4 waveform model including the full spin evolution and higher-order modes through the merger and ringdown.

\begin{figure}
	\centering
	\includegraphics[width=0.5\textwidth]{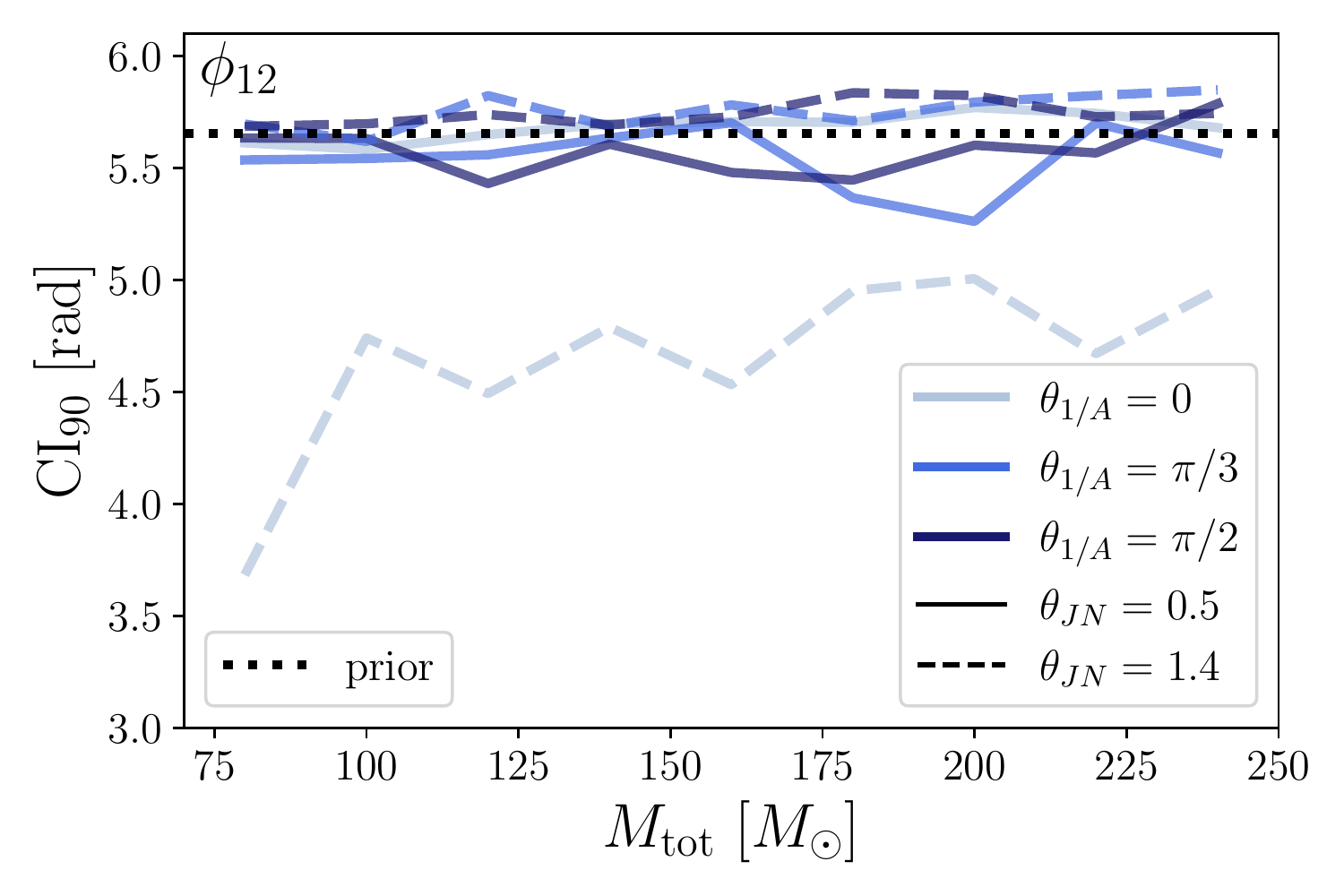}
	\caption{90\% credible interval for the $\phi_{12}$ posteriors obtained for the first set of simulations presented in Table~\ref{tab:injections}.}
	\label{fig:phi12_inj1}
\end{figure}

\begin{figure}
\centering
\includegraphics[width=0.5\textwidth]{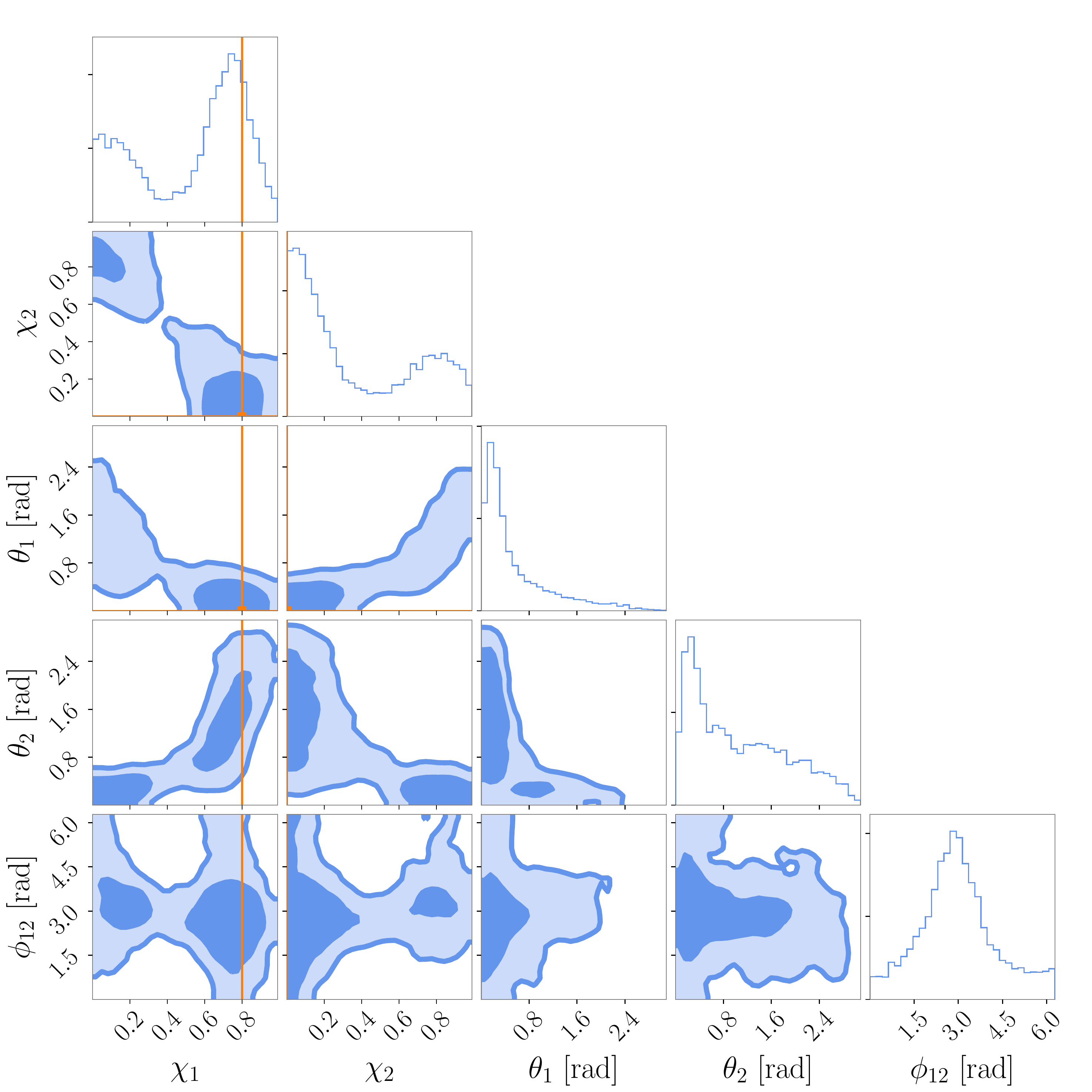}
\caption{Corner plot showing the posteriors for the component spin parameters
and $\phi_{12}$ for a simulated system with an aligned primary spin and
nonspinning secondary component. The orange lines show the true parameter
values, and hence are omitted for $\theta_{2}$ and $\phi_{12}$ as these
parameters do not have a physical meaning when only one object in the binary is
spinning.} 
\label{fig:phi12_corner}
\end{figure}

The 90\% credible intervals for the $\phi_{12}$ posteriors obtained for the equal-mass simulations presented in the previous section are shown in Fig.~\ref{fig:phi12_inj1}. For all the unequal-mass systems (not shown in Fig.~\ref{fig:phi12_inj1}) and all but one of the equal-mass configurations, the 90\% CI includes nearly the entire prior range. However, we find that the posterior can be better constrained for systems with aligned primary spin and equal mass when observed nearly edge-on (light blue dashed line) compared to all the other simulated configurations. A corner plot showing the posteriors for the spin parameters for one such binary where $\phi_{12}$ is well-measured is shown in Fig.~\ref{fig:phi12_corner}.
In this case, we recover a high primary spin magnitude with a small primary tilt angle and a small secondary spin magnitude with a large tilt angle, so the secondary spin is reconstructed to lie almost entirely in the orbital plane (even though the true value is $\vec{\chi}_B = 0$). 

If the perpendicular components of the spin vectors cancel, such that $\vec{S}_{1, \perp} + \vec{S}_{2, \perp} = 0$, then the total angular momentum lies entirely along the orbital angular momentum, so there is no precession. This can be achieved when $\phi_{12} = \pi$. 
Because the system is observed nearly edge-on, any modulations in the waveform due to orbital-plane precession would be more easily observable, so the lack of precession in this system means that the in-plane spin components should cancel, leading to the preference for $\phi_{12}=\pm \pi$. 
The lack of precession is not as clear when the system is observed nearly face-on, leading to a worse constraint on $\phi_{12}$. A similar constraint on $\phi_{12}$ is not obtained for the equivalent unequal-mass systems, since in this case the primary magnitude and tilt are very well-measured, but the secondary spin is relatively unconstrained. This leaves no degeneracy between the primary and secondary in-plane spin vectors that needs to accommodate the lack of precession in the signal by preferring $\phi_{12} = \pi$.

\begin{figure}
	\centering
	\includegraphics[width=0.5\textwidth]{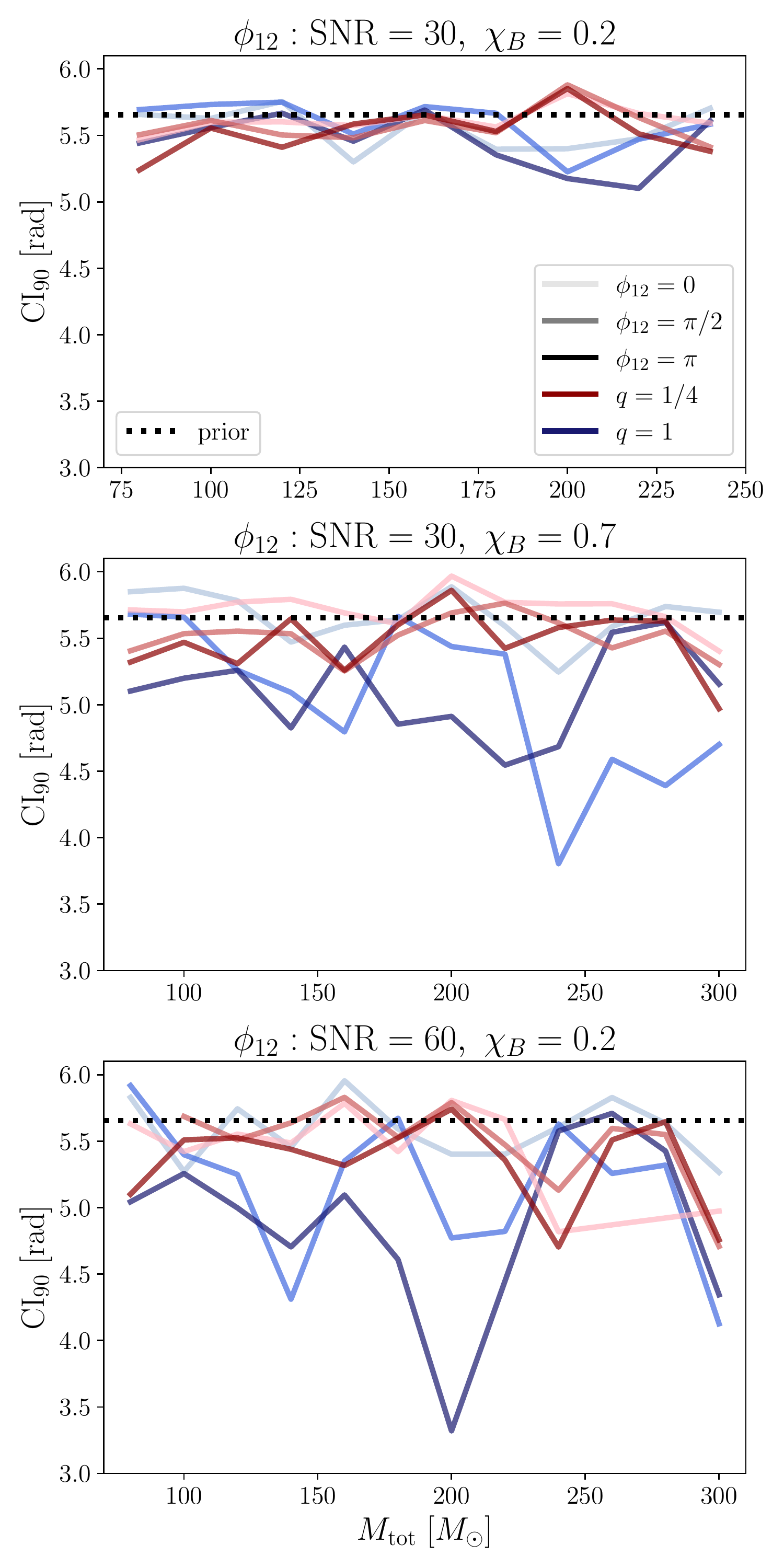}
	\caption{90\% credible intervals for the $\phi_{12}$ posteriors for Simulation Sets 2 (top; SNR=30, $\chi_{2}=0.2$), 3 (middle; SNR=30, $\chi_{2}=0.7$) and 4 (bottom; SNR=60, $\chi_{2}=0.2$). The other parameters of the simulated systems are given in Tables~\ref{tab:injections} and \ref{tab:same_params}.}
	\label{fig:phi12_inj234}
\end{figure}

Because $\phi_{12}$ does not have a physical meaning for systems with only one spinning component, we perform three additional sets of simulations, described by Simulation Sets 2--4 in Table~\ref{tab:injections}. In these new simulations, we try two different values of the secondary spin and two different values of the SNR, to verify whether higher-SNR events will improve the measurability of $\phi_{12}$. We also choose three different values of $\phi_{12}$, although the systems we simulate are \textit{not} necessarily in resonant configurations, which would require also choosing the spin tilt angles so that the value of $\phi_{12}$ is constant in time. The 90\% credible intervals for the $\phi_{12}$ posteriors for Simulation Sets 2, 3, and 4 are shown in Fig.~\ref{fig:phi12_inj234}.

For the systems with $\mathrm{SNR}=30$ and low secondary spin, $\chi_{2}=0.2$, shown in the top panel of Fig.~\ref{fig:phi12_inj234}, $\phi_{12}$ is unconstrained relative to the prior regardless of the value of $\phi_{12}$ or the mass ratio of the system across all the total masses we analyzed. While the measurements  improve slightly for equal-mass systems when the secondary spin is increased to $\chi_{2}=0.7$ in the middle panel and when the SNR is increased to 60 in the bottom panel, $\phi_{12}$ remains generally poorly constrained at 25 Hz for most systems we analyze.

\begin{figure}
	\centering
	\includegraphics[width=0.5\textwidth]{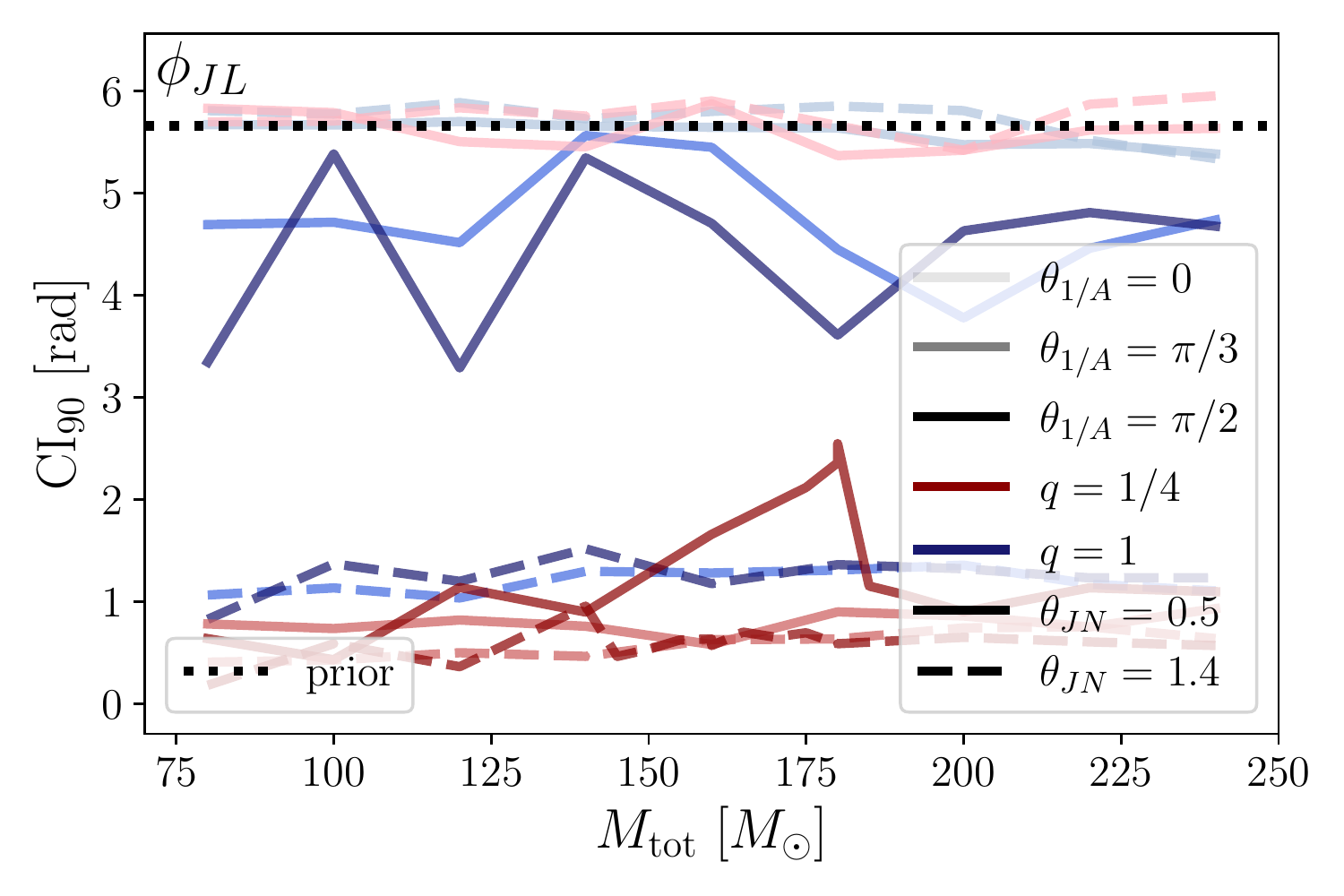}
	\caption{90\% credible interval for the $\phi_{JL}$ posteriors obtained for the first simulation set after wrapping the posterior samples around $\phi_{JL}=2\pi$.}
	\label{fig:phi_JL}
\end{figure}

The remaining spin degree of freedom is the azimuthal phase of the total angular momentum, $\vec{L}$, in its precession cone around the total angular momentum, $\vec{J}$, called $\phi_{JL}$ (see right panel of Fig~\ref{fig:schematic}).
Along with the other five spin parameters and the binary mass ratio, this angle determines the recoil velocity and direction of the remnant black hole, which can have implications for the rate of hierarchical mergers in dynamical environments like globular clusters~\cite{Varma:2020nbm}. Because the true value of $\phi_{JL}$ used in our simulations is at the edge of the prior (see Table~\ref{tab:same_params}), the posterior probability density is bimodal with peaks at $\phi_{JL}=0, 2\pi$. In order to meaningfully calculate the width of the 90\% credible interval, we map the posterior samples with $\phi_{JL}<\pi$ to $\phi_{JL}+2\pi$. The resulting 90\% credible interval for $\phi_{JL}$ as a function of the total mass of the simulated system is shown in Fig.~\ref{fig:phi_JL}.

The worst constraint on $\phi_{JL}$ comes from systems with aligned primary spin, consistent with the fact that this parameter is not physically meaningful for aligned-spin systems where the total and orbital angular momentum vectors point in the same direction. The measurement of $\phi_{JL}$ improves for system with unequal mass ratios compared to those of equal mass and for systems observed nearly edge-on compared to those observed nearly face-on. Systems with unequal mass ratios have fewer degeneracies and enhanced higher-order mode emission, making the presence of precession---and hence an offset between the total and orbital angular momenta---easier to measure. This effect is similarly easier to discern for systems viewed nearly edge-on, when the effects of precession are enhanced. The primary tilt angle does not have a significant effect on the width of the $\phi_{JL}$ posterior for systems with the same mass ratio and inclination. We conclude that, while $\phi_{JL}$ can be well-measured for systems where the effects of precession are apparent, $\phi_{12}$ is generally poorly constrained even for highly spinning systems and those with high SNR, but leave further investigation of the measurability of $\phi_{12}$ for systems locked in spin-orbit resonance to future work~\cite{Varma:2021csh}.

\section{Conclusions}
We have presented a systematic analysis of the measurability of spin in heavy binary black hole sources including both aligned and precessing spins. We used the NRSur7dq4 waveform~\cite{Varma:2019csw}, which models the effects of all six spin degrees of freedom on the signal, as well as higher-order modes up to $\ell \leq 4$. Motivated by the informative measurement of $\chi_{p}$ for GW190521~\cite{Abbott:2020tfl}, we performed several simulations of systems with parameters consistent with the GW190521 posterior, but were unable to recover an informative measurement of $\chi_{p}$ until we used the parameters of the maximum-likelihood posterior sample as the true values for our simulated system, indicating that the correlations between parameters and the exact six-dimensional spin configuration have a significant effect on the measurability of $\chi_{p}$. For this maximum-likelihood system, we find that the spin measurement hinges on the merger and ringdown part of the signal.

We extended our study to generic configurations to investigate how the bias and precision of the spin posteriors scale with the total mass of the system. We find that the spin magnitude of the highest-spinning black hole~\cite{Biscoveanu:2020are}, $\chi_{A}$, can be accurately constrained to a 90\% credible interval width of $\lesssim 0.5$ for systems with aligned spin or unequal masses (compared to a prior width of $0.89$). 
Because of the degeneracies in the system, the spin of the lowest-spinning black hole, $\chi_{B}$, is marginally better-constrained for equal-mass systems, but the posteriors for this parameter are uninformative for unequal-mass systems. The constraint on $\chi_{B}$ worsens as the total mass of the system increases for equal-mass systems, and the same is true for $\chi_{A}$ for the unequal-mass systems. No significant trend is observed for the width of the $\chi_{A}$ posterior as a function of mass for equal-mass systems.

The tilt of the highest-spinning black hole, $\theta_{A}$, can be measured with a posterior 90\% CI width of $\lesssim 1.5$ rad, and is in general better-measured for unequal-mass systems~\cite{Cho:2012ed, OShaughnessy:2014shr, Vitale:2014mka, Pankow:2016udj, Pratten:2020igi}. The corresponding prior width for $\theta_{A}$ is $2.21$ rad. Both the spin magnitude and tilt of the highest-spinning black hole, $\chi_{A}$ and $\theta_{A}$, are more accurately and precisely measured for aligned-spin systems, since the absence of precession is informative. We also find a generic improvement in the posterior bias and width for systems viewed nearly edge-on compared to those observed face-on~\cite{Vitale:2014mka}, as the observer can see both sides of the orbital plane as it oscillates for some orientations when $\cos{\theta_{JL}} < \sin{\theta_{JN}}$, enhancing the visibility of precession for a fixed SNR.

Interestingly, we find that the effective aligned spin, \chieff, is in general measured to a posterior 90\% credible interval width of $\lesssim 0.35$ for all the high-mass systems we simulate (prior width of $0.84$), despite the limited number of inspiral cycles available in the analysis.
While the width increases as a function of mass for most of the unequal-mass systems, it actually narrows at the highest masses for equal-mass systems. On the other hand, $\chi_{p}$ is not well-constrained for equal-mass systems, but is in general measured to a posterior 90\% credible interval width of $\lesssim 0.5$ for unequal-mass systems (compared to a prior width of $0.75$). As was the case for the component spins, the effective spins are more narrowly constrained for aligned-spin systems, although the median of the $\chi_{p}$ posterior is biased away from the true value by ${\sim} 0.4$ for equal-mass systems regardless of the primary tilt angle.

The trends in the width of the 90\% credible interval of the $\chi_{p}$ posterior as a function of mass appear to exhibit significant outliers for unequal-mass systems with in-plane primary tilt at ${\sim} 160~M_{\odot}$ for systems observed nearly edge-on and at ${\sim} 180~M_{\odot}$ when observed nearly face-on. These apparent outliers are due to the variation in the exact spin configuration at a fixed dimensionless reference frequency. Changing the total mass of the simulated systems but keeping the reference frequency fixed to $25~\mathrm{Hz}$ results in variations in the true simulated primary tilt angle at fixed dimensionless reference frequency of the order of a tenth of a radian. When repeating the analysis with a fixed dimensionless reference frequency, these apparent outliers disappear, reinforcing the previous indication that small changes in the exact spin configuration can have a significant effect on the measurability of the spin parameters. Reanalyzing the systems with a fixed reference frequency of $25~\mathrm{Hz}$ with two different waveform models, IMRPhenomPv2~\cite{Hannam:2013oca, Husa:2015iqa, Khan:2015jqa} and IMRPhenomXPHM~\cite{Pratten:2020ceb, Pratten:2020fqn, Garcia-Quiros:2020qpx}, led to different results, as shown in detail in Appendix~\ref{sec:waveforms}.

We also investigate the measurability of the azimuthal spin angles that complete the six spin degrees of freedom. The $\phi_{12}$ angle can encode information about the system's formation history via spin-orbit resonances~\cite{Schnittman:2004vq, Gerosa:2013laa, Gerosa:2014kta}, and both $\phi_{12}$ and $\phi_{JL}$ affect the recoil kick of the remnant black hole~\cite{Varma:2020nbm}. While $\phi_{JL}$ is generally measured to a posterior 90\% credible interval width of $\lesssim 1.5$ (prior width $5.65$) for systems with some misalignment with unequal masses or viewed nearly edge-on, $\phi_{12}$ is generally unconstrained regardless of the SNR or the spin magnitude of the secondary for the configurations we simulated. The lack of constraint for $\phi_{12}$ could potentially be explained by the choice of reference frequency, as Ref.~\cite{Varma:2021csh} recently showed that measuring the spins close to merger improves the constraint on this parameter. We do find, however, that some aligned-spin systems preferentially recover posteriors peaked at $\phi_{12} = \pm \pi$, since this configuration allows the effects of precession, which are absent in the signal, to cancel in the posterior.

To summarize, we conclude that:
\begin{enumerate}
\item It is possible to capture the effects of precession via the posterior on $\chi_{p}$ for highly-precessing, heavy systems with moderate SNR like GW190521, but the measurability of $\chi_{p}$ depends on the exact six-dimensional spin configuration.
\item For a highly precessing system including the parameter correlations that led to the informative $\chi_{p}$ measurement for GW190521, the inference of the component spins and hence $\chi_{p}$ depends strongly on the merger and ringdown parts of the signal, rather than the inspiral.
\item Consistent with previous studies, the spin of the highest-spinning black hole and $\chi_{p}$ are better measured for systems with unequal masses compared to those with equal masses. All spin parameters are better measured when the system is observed at higher inclination angles for systems with total masses in the range $80-240~M_{\odot}$.
\item The spin parameters are better constrained for systems with aligned spins compared to those in generically precessing configurations, since the lack of precession leaves a characteristic imprint on the waveform in these cases. Even when allowing for precession in the recovery of these systems, the lack of obvious signs of precession in the data leads to improved constraints on the spin parameters.
\item \chieff is well-measured for all the masses we considered, with a 90\% posterior credible interval width of $\lesssim 0.35$, demonstrating that the effective spin can be inferred in the absence of a prominent inspiral. For comparison, the prior width for \chieff is $0.84$. Furthermore, the constraint improves as the mass increases for some configurations, particularly those with equal mass. 
\item Variations of even a tenth of a radian in the primary tilt angle at a fixed dimensionless reference frequency can lead to significant differences in the ability to accurately and precisely measure the spins of a particular system.
\item The azimuthal angle between the two component spin vectors is in general unconstrained at a fixed reference frequency of $25~\mathrm{Hz}$, even for systems detected at high SNR. However it can be recovered with a preference towards $\phi_{12}=\pm \pi$ for some aligned-spin systems, even though the true value of this parameter is not physically meaningful in these cases.
\end{enumerate}

Our study raises several questions, particularly about the optimal configuration for measuring the effects of precession. While we find that the constraint on $\chi_{p}$ varies significantly depending on the exact six-dimensional spin configuration of the system, a more detailed exploration of the parameter space would be necessary to reveal which configurations lead to the best measurement of $\chi_{p}$. We find that the true value of \chieff or $\chi_p$ alone is not enough to explain these variations. 
Furthermore, despite the fact that we find that the spin information is primarily driven by the merger and ringdown parts of the signal for the GW190521 maximum-likelihood simulation, the uncertainty in the component spin parameters continues to increase as the total mass of the system increases, even for those systems simulated at a fixed dimensionless reference frequency. 
This suggests that losing inspiral cycles leads to a loss of spin information, but further investigations are required to discern the effects of the inspiral vs. postmerger parts of the signal on the measurability of spin in generic systems.
Finally, while we find that $\phi_{12}$ is unconstrained for most systems explored in this work, these systems are not necessarily locked in spin-orbit resonances. We leave the investigation of the measurability of $\phi_{12}$ for resonant systems to future work.

\begin{acknowledgments}
The authors would like to thank Isobel Romero-Shaw for sharing the details of the simulations performed in \cite{Romero-Shaw:2020thy} with us and for comments on the manuscript. They also thank Davide Gerosa, Carl-Johan Haster, Aaron Zimmerman, Geraint Pratten, Sascha Husa, Maria Haney, Charlie Hoy, Rhys Green, Patricia Schmidt, Ajit Mehta, and Christopher Berry for useful discussions.
S.B., M.I. and S.V.\ acknowledge support of the National Science Foundation and the LIGO Laboratory.
LIGO was constructed by the California Institute of Technology and
Massachusetts Institute of Technology with funding from the National
Science Foundation and operates under cooperative agreement PHY-1764464.
This research has made use of data, software, and/or web tools obtained from the Gravitational Wave Open Science Center (\href{https://www.gw-openscience.org}{https://www.gw-openscience.org}), a service of LIGO Laboratory, the LIGO Scientific Collaboration and the Virgo Collaboration. Virgo is funded by the French Centre National de Recherche Scientifique (CNRS), the Italian Istituto Nazionale della Fisica Nucleare (INFN), and the Dutch Nikhef, with contributions by Polish and Hungarian institutes.
S.B. is also supported by the Paul and Daisy Soros Fellowship for New Americans and the NSF Graduate Research Fellowship under Grant No. DGE-1122374.
M.I.\ is supported by NASA through the NASA Hubble Fellowship
grant No.\ HST-HF2-51410.001-A awarded by the Space Telescope
Science Institute, which is operated by the Association of Universities
for Research in Astronomy, Inc., for NASA, under contract NAS5-26555.
V.V.\ is supported by a Klarman Fellowship at Cornell.
The authors are grateful for computational resources provided by the LIGO Lab and supported by NSF Grants PHY-0757058 and PHY-0823459.
This paper carries LIGO document number LIGO-P2100204.
\end{acknowledgments}

\appendix
\section{GW190521 maximum-likelihood simulation}
\label{ap:GW190521}
\begin{figure*}
\centering
\includegraphics[width=0.95\columnwidth]{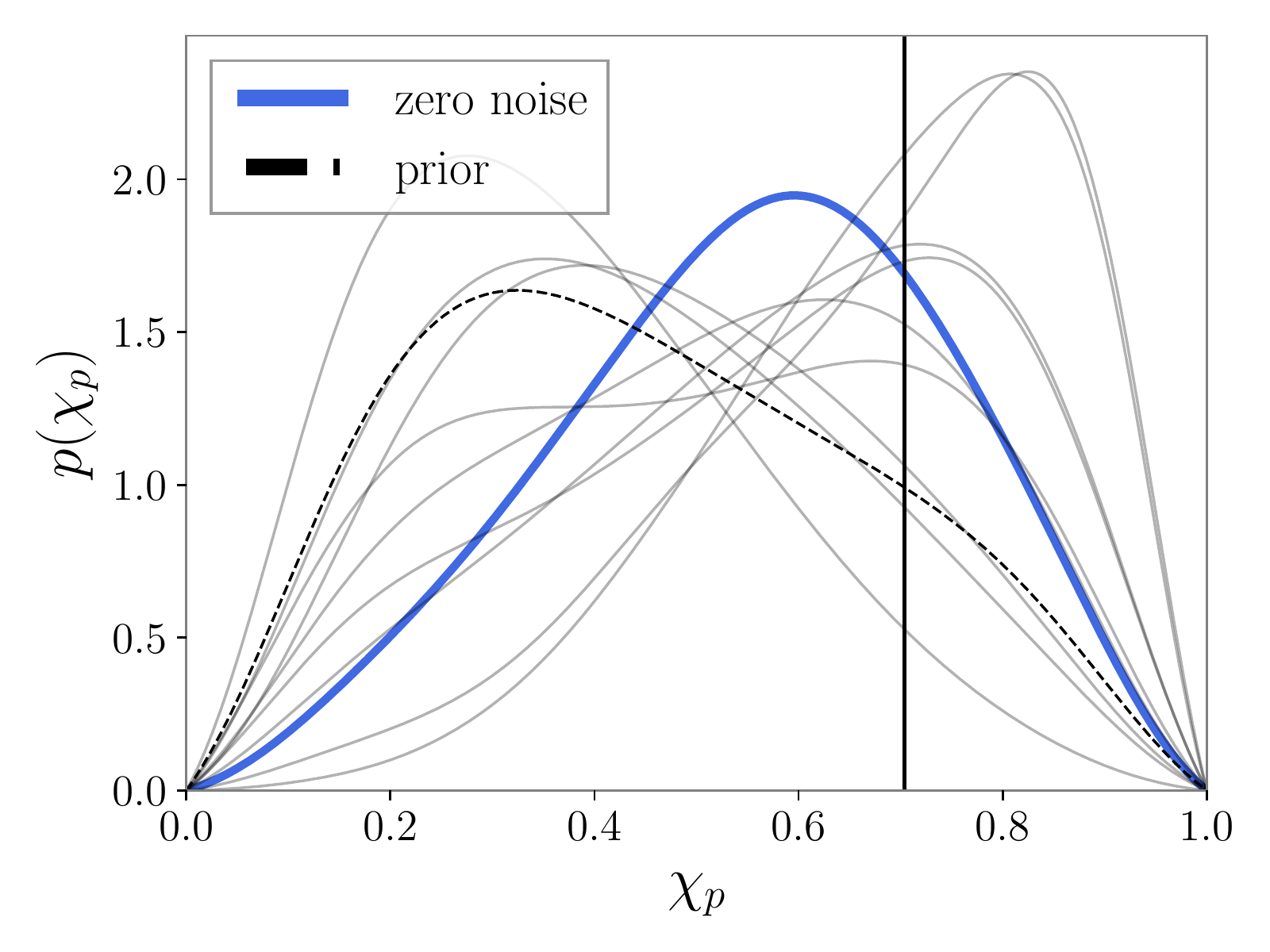}
\includegraphics[width=0.95\columnwidth]{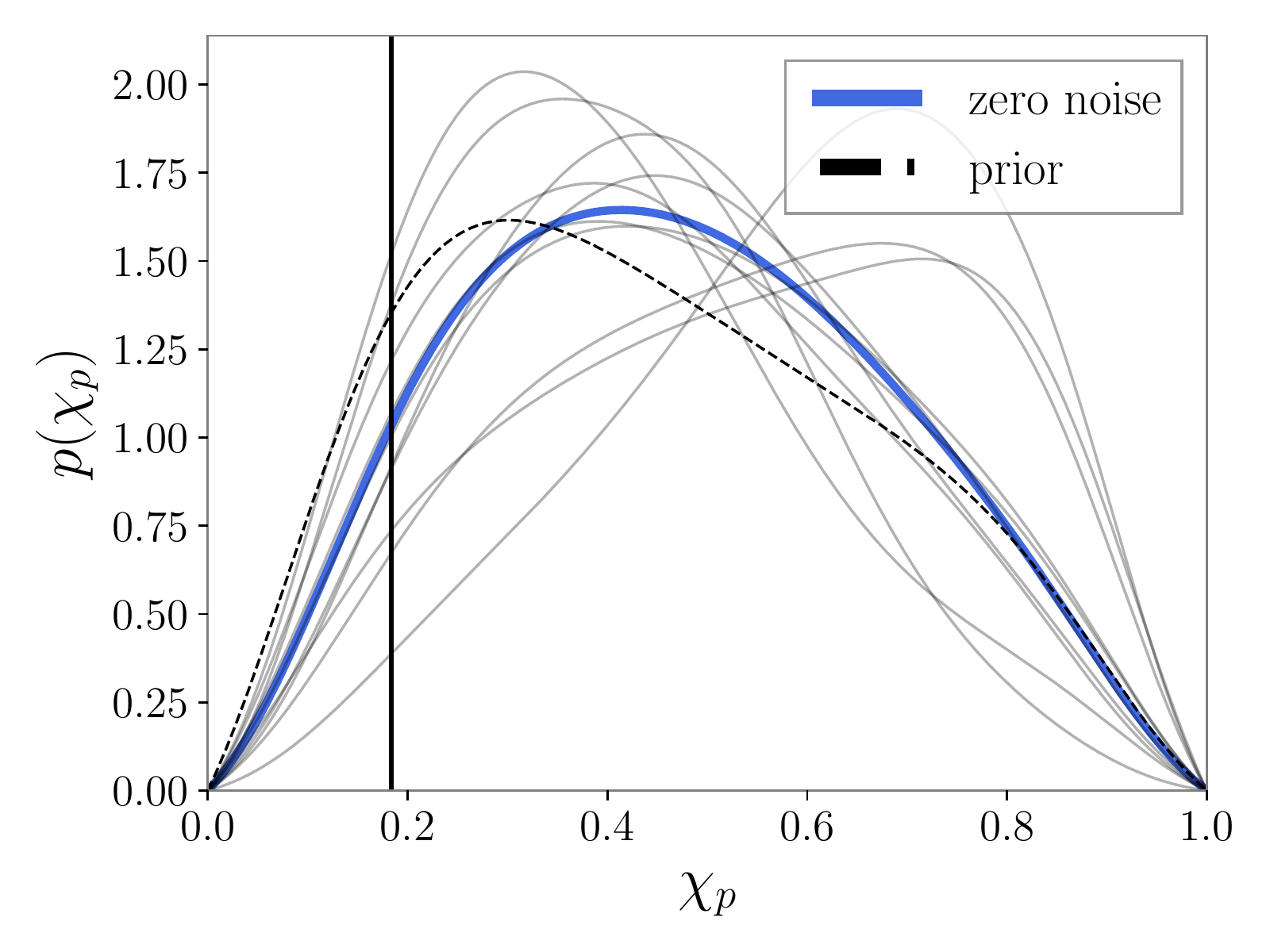}
\caption{\textit{Left:} Probability density for $\chi_{p}$ for the GW190521 maximum-likelihood
simulation added to ten different segments of real detector noise from O3. \textit{Right:} Probability density for $\chi_{p}$ for the simulated system with low-$\chi_{p}$ drawn from the GW190521 posterior added to ten different realizations of Gaussian noise colored by the PSDs calculated for the data segment containing GW190521. The thick, dark blue line shows the result of the corresponding simulation without Gaussian
noise for comparison.
}
\label{fig:GW190521_real}
\end{figure*}

For the simulations where the GW190521 maximum-likelihood system is added to real detector noise from O3, we choose ten distinct $4~\mathrm{s}$ data segments beginning $8~\mathrm{s}$ after the coalescence times of ten of the higher-mass BBH systems included in GWTC-2 and assume the data in each segment is well-characterized by the PSDs computed for each of those events (publicly available at \cite{data_release_GWTC2}). The events and true coalescence times we use are detailed in Table~\ref{tab:events_times}. For these analyses, we use a bandwidth of $16-256~\mathrm{Hz}$ and a sampling rate of $512~\mathrm{Hz}$.
We only include the data from the Hanford and Livingston detectors in this set of simulations, since Virgo data was not available for all of our chosen events. Because we allow the PSDs to vary, we adjust the distances of the ten simulated systems to keep the network matched filter SNR fixed to 15 for consistency with the Gaussian noise tests. The results of the real noise simulations are shown in the left panel of Fig.~\ref{fig:GW190521_real}. The variation in the $\chi_{p}$ posterior is similar to that observed in Fig.~\ref{fig:GW190521_high}.

\begin{table}
\begin{ruledtabular}
\begin{tabular}{l l}
    Event & Time \\
    \midrule
    GW190413\_134308 & 1239198214.76 \\
    GW190421\_213856 & 1239917962.27 \\
    GW190503\_185404  & 1240944870.30  \\
    GW190517\_055101  & 1242107487.84 \\
    GW190527\_092055 & 1242984081.81 \\
    GW190602\_175927 & 1243533593.11 \\
    GW190706\_222641 & 1246487227.35 \\
    GW190719\_215514 & 1247608540.95 \\
    GW190909\_114149 & 1252064535.73 \\
    GW190915\_235702 & 1252627048.71 \\
\end{tabular}
\end{ruledtabular}
\caption{Events from O3 whose PSDs were used and the corresponding coalescence times at which we added the GW190521 maximum-likelihood simulation into real detector data.}
\label{tab:events_times}
\end{table}

We also repeat the experiment using Gaussian noise colored by the PSDs calculated for the data segment containing GW190521 but instead add a system with parameters drawn from the GW190521 posterior with a low value of $\chi_{p}$. The resulting $\chi_{p}$ posteriors are shown in the right panel of Fig.~\ref{fig:GW190521_real}, and the true values of the parameters are given in Table~\ref{tab:GW190521_low}. Although the true value of $\chi_{p}=0.18$, these posteriors much more closely resemble those obtained for the three initial highly-precessing systems shown in Fig.~\ref{fig:chip_corner}. They are less informative and peak at lower values of $\chi_{p}$, driven by the prior. Although there is one noise realization that results in a posterior peaking at high $\chi_{p}$, this is consistent with Gaussianity. Taken together, the results for all three experiments using draws from the posterior of GW190521 as the true simulation parameters indicate that while it is not possible to rule out the hypothesis that the informative posterior on $\chi_{p}$ for GW190521 is driven by noise fluctuations at the time of the event, it is more likely due to the exact spin configuration that leads to $\chi_{p}$ being easier to measure.

\begin{table}
\begin{ruledtabular}
\begin{tabular}{l l l}
    Parameter &Symbol & Value\\
    \midrule
    Mass ratio & $q$ & 0.76 \\
    Detector-frame total mass & $M_{\mathrm{tot}}$ & $278.63~M_{\odot}$ \\
    Primary spin magnitude & $\chi_{1}$  & 0.18  \\
    Secondary spin magnitude & $\chi_{2}$  & 0.52 \\
    Primary tilt\footnote{All angles in radians.} & $\theta_{1}$ & 1.62 \\
    Secondary tilt & $\theta_{2}$ & 0.19 \\
    Azimuthal inter-spin angle & $\phi_{12}$ & 5.99 \\
    Azimuthal precession cone angle & $\phi_{JL}$ & 4.73 \\
    Coalescence phase & $\phi$ & 5.92 \\
    Polarization angle & $\psi$ & 2.48 \\
    Coalescence GPS time & $t_{c}$ & 1242442967.41 s\\
    Right ascension & $\alpha$ & 0.13 \\
    Declination & $\delta$ & -1.17 \\
    Luminosity distance & $d_{L}$ & 4735.54 Mpc \\
    Inclination angle & $\theta_{JN}$ & 0.56\\
\end{tabular}
\end{ruledtabular}
\caption{Parameter values drawn from the GW190521 posterior for the simulation with low $\chi_{p}$.}
\label{tab:GW190521_low}
\end{table}

\section{Mass measurements for simulated single-spin systems}
\label{ap:masses}
\begin{figure*}
	\centering
	\includegraphics[width=\textwidth]{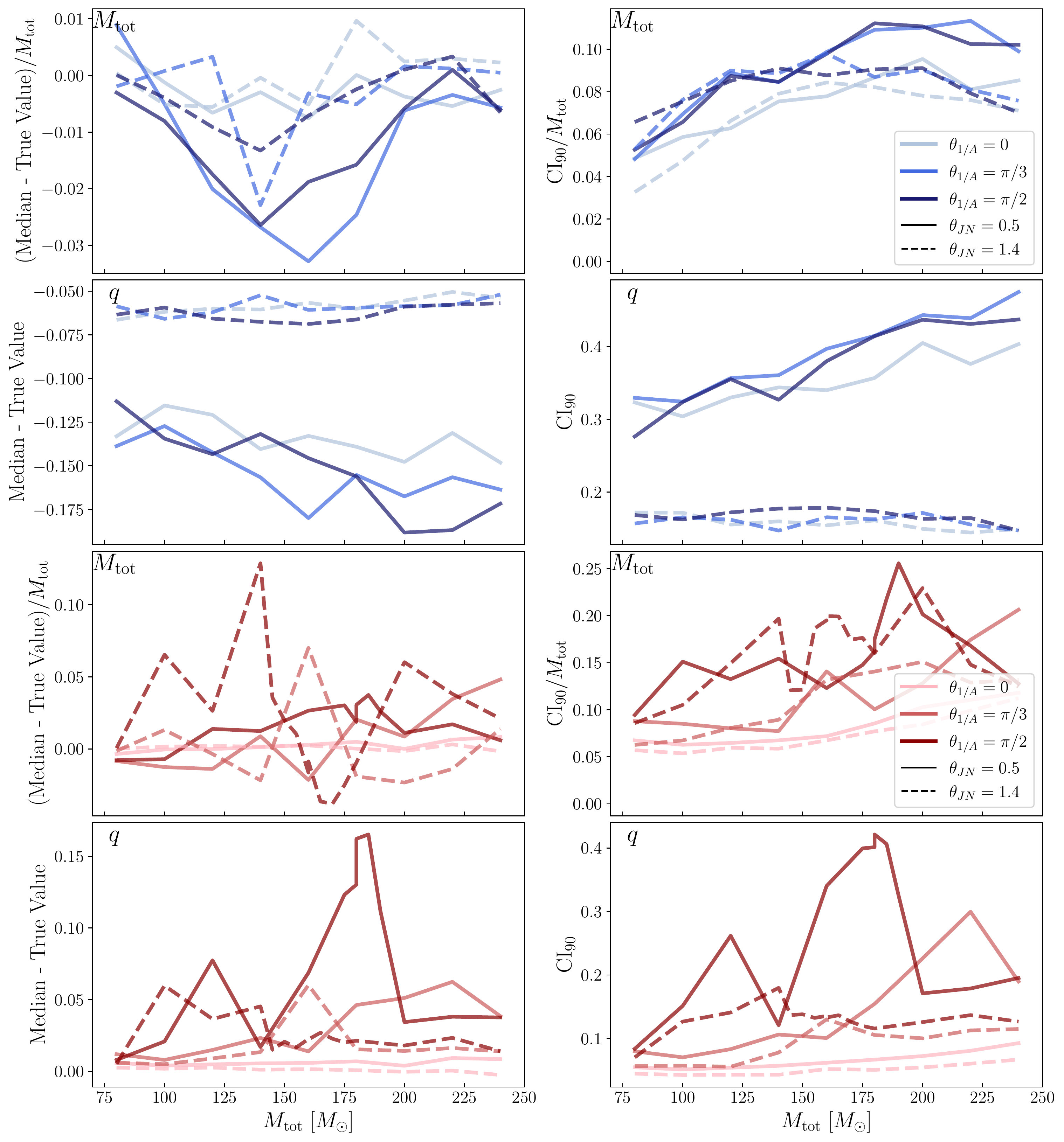}\qquad
	\caption{The bias and width of the total mass and mass ratio posteriors as a function of the total mass of the system for the first simulation set described in Table~\ref{tab:injections}. The bias is defined as the difference between the median of the posterior and the true value, while the width is the symmetric 90\% credible interval. For the total mass, these quantities are normalized by the value of the total mass of each system so that they are dimensionless. Equal-mass systems are shown in blue, while unequal-mass systems are shown in red. Systems observed nearly face-on are shown with a solid line, while those observed nearly edge-on are represented by a dotted line. The shading of the line corresponds to the tilt of the primary spin.}
	\label{fig:mass_bias}
\end{figure*}

The bias and uncertainty in the total mass and mass ratio posteriors for the simulations discussed in detail in Section~\ref{sec:bias_vs_mass} are shown in  Fig.~\ref{fig:mass_bias}. For the total mass, we normalize the bias and credible interval width by the value of the total mass of each system so that these quantities are dimensionless, consistent with the other parameters. Several trends observed for the spin parameters are also present in the measurements of the mass parameters. The width of the 90\% CI increases with increasing total mass for the total mass of the system for all configurations, and for the mass ratio for equal-mass systems observed nearly face-on (solid blue lines). Both mass parameters are also generally more accurately and precisely recovered for aligned-spin systems (lightest line color) compared to those with some primary spin tilt. While the total mass constraints are similar for systems observed nearly edge-on and nearly face-on, the mass ratio measurement improves considerably for systems observed nearly edge-on in terms of both the bias and the posterior width, particularly for equal-mass systems. This can be explained due to the enhanced observability of higher-order-mode emission at higher inclinations. The equal-mass systems are always biased towards lower values of $q$ by definition, since the median of the posterior will always fall below the true value of $q=1$, which is at the upper edge of the prior.

The outliers discussed in Section~\ref{sec:outliers} for the spin parameters are also present in the mass parameters. The location of the outliers in total mass and mass ratio for the system observed nearly face-on (dark red solid line) matches with that of the outliers in the spin parameters for this system. However, the outlier is absent in the mass ratio posterior for the system observed nearly edge-on, and the peak in the width of the 90\% credible interval for the total mass posterior shifts towards higher masses than the corresponding peak in the posteriors of the spin parameters. Comparing with the results for the systems analyzed at a fixed dimensionless reference frequency shown in Fig.~\ref{fig:fixed_frefMtot}, we conclude that these outliers in the mass parameters are also caused by differences in the spin angles of the underlying systems due to using a fixed reference frequency of 25 Hz for the analysis.

\section{Waveform systematics}
\label{sec:waveforms}
In order to verify whether the observed outliers described in Section~\ref{sec:outliers} are unique to the NRSur7dq4 waveform model, we repeat the original simulations with a fixed reference frequency of  $25~\mathrm{Hz}$ for the systems demonstrating the outlier behavior with two additional waveform models, IMRPhenomPv2~\cite{Hannam:2013oca, Husa:2015iqa, Khan:2015jqa}  and IMRPhenomXPHM~\cite{Pratten:2020ceb, Pratten:2020fqn, Garcia-Quiros:2020qpx}. These are both frequency-domain models, in contrast to the numerical relativity surrogate, which is a time-domain model. IMRPhenomPv2 includes only the dominant $\ell = |m| =2$ multipoles, while IMRPhenomXPHM includes a subset of the higher-order modes included by the surrogate. Both of these phenomenological models incorporate the effects of precessing spins by approximating the full waveform as an underlying non-precessing waveform in the co-precessing frame that gets ``twisted up''  into the inertial frame via a rotation encoding the evolution of the orbital plane~\cite{Schmidt:2012rh}.
IMRPhenomPv2 uses the next-to-next-to-leading (NNLO) order, single spin, post-Newtonian approach described in \cite{Schmidt:2014iyl}, where $\vec{\chi}_{2}=0$, and only includes spin-orbit (not spin-spin) interactions. IMRPhenomXPHM adds the option of another implementation based on the multiple-scale analysis (MSA) introduced in \cite{Chatziioannou:2017tdw} and previously applied in \cite{Khan:2018fmp, Khan:2019kot}, which enables the inclusion of double-spin effects and includes some contributions from spin-spin coupling. We use the MSA prescription for the Euler angles for our analysis, falling back to the NNLO presciption if the MSA system fails to initialize.  

\begin{figure*}
	\centering
	\includegraphics[width=\textwidth]{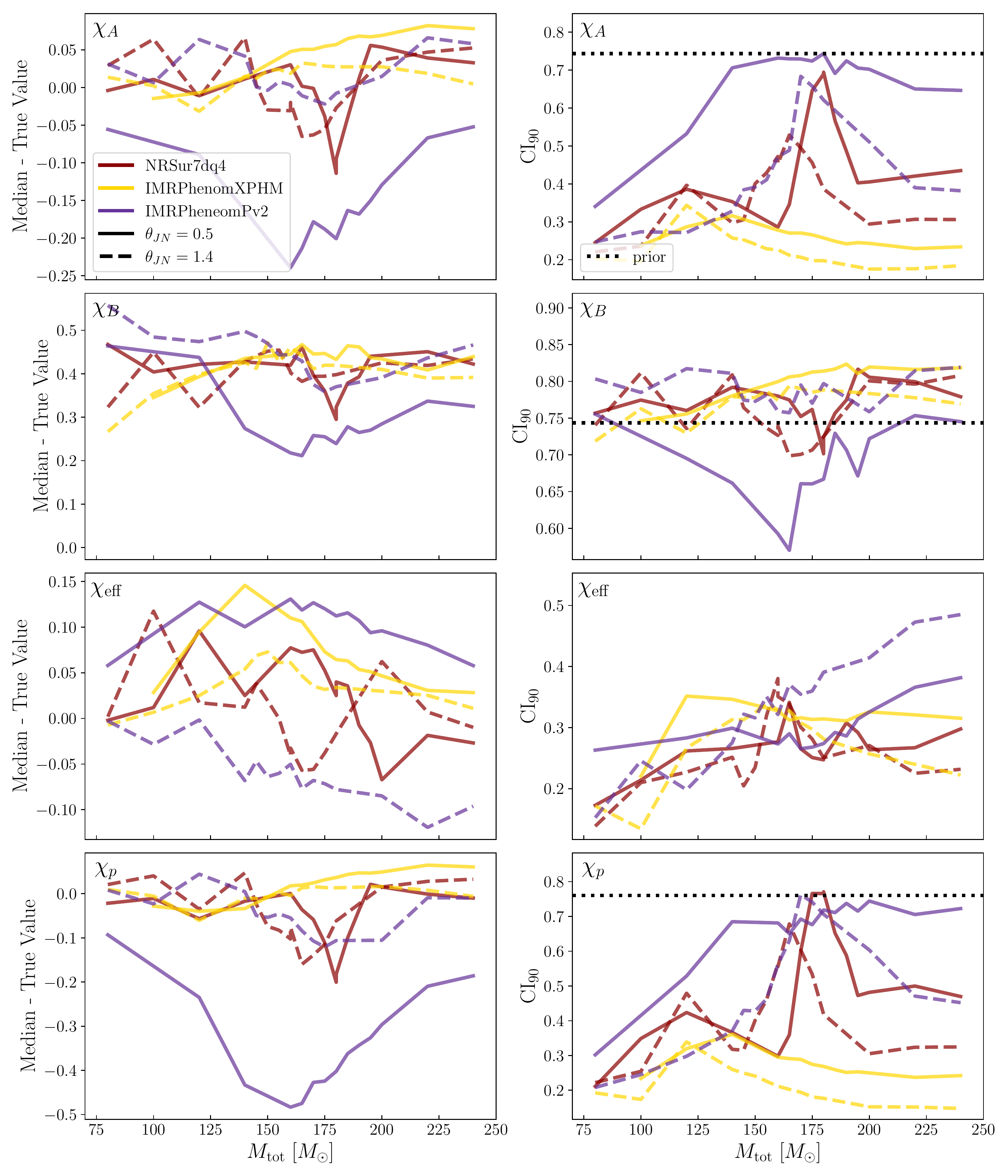}
	\caption{Bias and width of the 90\% credible interval of the posteriors for component spin magnitudes and effective spins as a function of the true total mass for the original systems with outliers analyzed with three different waveform models: NRSur7dq4 (red), IMRPhenomXPHM (gold), and IMRPhenomPv2 (purple). The systems viewed nearly edge-on are indicated with dashed lines and those observed nearly face-on with solid lines. The dotted black lines show the width of the 90\% CI for the prior for all parameters except \chieff, for which the prior width of 0.84 is much larger than the posterior widths.}
	\label{fig:approx_comp_spin}
\end{figure*}

\begin{figure*}
	\centering
	\includegraphics[width=\textwidth]{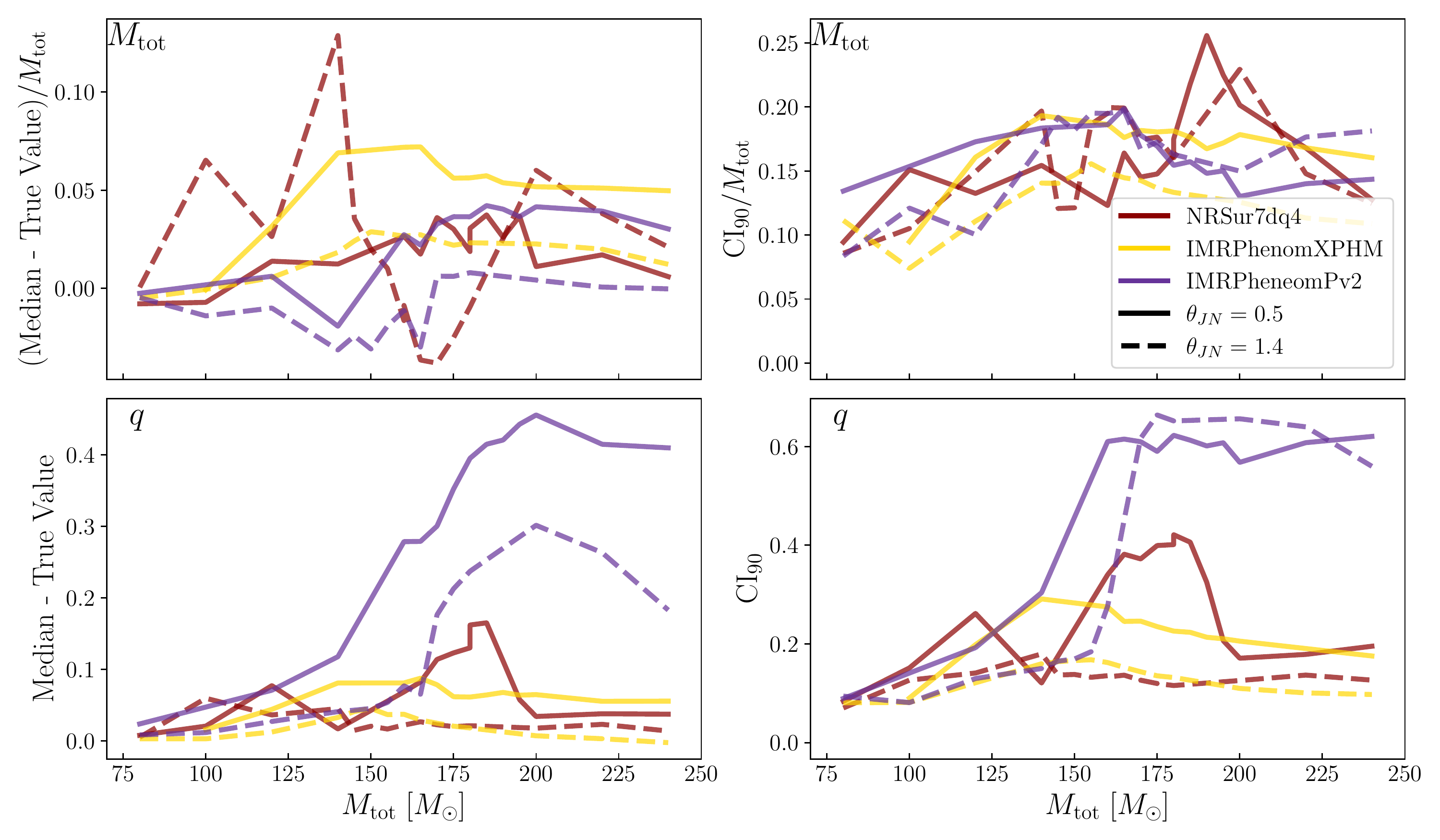}
	\caption{Bias and width of the 90\% credible interval of the posteriors for the total mass and mass ratio as a function of the true total mass for the original systems with outliers analyzed with three different waveform models: NRSur7dq4 (red), IMRPhenomXPHM (gold), and IMRPhenomPv2 (purple). The systems viewed nearly edge-on are indicated with dashed lines and those observed nearly face-on with solid lines.}
	\label{fig:approx_comp_masses}
\end{figure*}

A comparison of the accuracy and precision of the spin and mass measurements for the three different waveform models is shown in Figs.~\ref{fig:approx_comp_spin}--\ref{fig:approx_comp_masses}. We use the same waveform model for both the simulation and recovery of these systems, so the purpose of this study is not to determine which model is the best at recovering the parameters of the same ``true'' underlying system. Rather, we want to determine if the trends we observe in the measurability of spins for high-mass systems are universal, particularly for the dramatic increase in the width of the 90\% credible interval observed at outlier masses. We find that these three waveform models produce different results.

IMRPhenomPv2 yields the largest bias and 90\% credible region for the primary spin parameters for the nearly face-on systems (purple solid lines in Fig.~\ref{fig:approx_comp_spin}). It systematically recovers values of $\chi_{p}$ that are smaller than the true value for systems with this inclination angle, and the $\chi_{p}$ posterior is basically unconstrained for total masses of $M_{\mathrm{tot}}\gtrsim 140~M_{\odot}$. The IMRPhenonPv2 results for the nearly edge-on systems (purple dashed lines) are closer to those obtained with the waveforms including higher-order modes, although \chieff is systematically biased towards lower values, and the width of the 90\% credible interval for \chieff is bigger by $\sim0.2$ compared to IMRPhenomXPHM and NRSur7dq4. The trends in the mass ratio posteriors are particularly striking for IMRPhenomPv2, where both the bias and the precision get much worse for total masses $M_{\mathrm{tot}}\gtrsim 140~M_{\odot}$. These results emphasize the improvement in the measurability of spin for precessing systems when viewed nearly edge-on compared to face-on and the importance of higher-order modes for such heavy systems with unequal mass ratios, the presence of which helps to place significantly more stringent constraints on the mass ratio of the system.

While IMRPhenomXPHM and NRSur7dq4 demonstrate qualitatively similar behavior for the bias and posterior width at low masses, $M_{\mathrm{tot}}\lesssim 160~M_{\odot}$, particularly for $\chi_{A}$ and $\chi_{p}$, the two disagree at the highest masses in the range of the outliers originally observed in the NRSur7dq4 results, which are not present with IMRPhenomXPHM. The posterior width has a smaller peak at around $M_{\mathrm{tot}}\approx 140~M_{\odot}$ for $\chi_{A}, \chi_{p}, q$, corresponding to a similar feature in the NRSur7dq4 results, but then continues to decrease gradually all the way up to the highest masses. A similar small peak is observed in the bias for \chieff, but no other parameters exhibit significant features in the bias. The results are qualitatively similar for systems observed at both inclinations. The significant differences in the results between these two waveform models that include similar physical effects indicate that the more sophisticated spin treatment of the NRSur7dq4 model could have an important impact on the measurability of spin for high-mass, precessing systems. Furthermore, IMRPhenomXPHM is not tuned to precessing NR simulations.  Since NR modifications are more important around the merger, which contributes a larger fraction of the SNR at higher masses, IMRPhenomXPHM is not as reliable as NRSur7dq4 in this regime.
In this case, the improvement in the spin constraints seen for IMRPhenomXPHM is artificial since the waveform is agnostic to the physics it fails to capture for the highest-mass systems, and we are using the same model for simulation and recovery.

\bibliography{spins}

\end{document}